\documentclass[preprint,3p,times,twocolumn,authoryear]{elsarticle}

\usepackage{amssymb}
\usepackage{amsmath}

\journal{Astronomy and Computing}

\DeclareUnicodeCharacter{2500}{--}

\usepackage{booktabs}
\usepackage{cuted}                      
\usepackage[inline]{enumitem}
\usepackage{float}
\usepackage[T1]{fontenc}
\usepackage{hyperref}
\usepackage{glossaries}
\usepackage{lineno}
\usepackage{listings, multicol}         
\usepackage{siunitx}
\usepackage{subfig}                     
\usepackage{threeparttable}
\usepackage{ulem}
\usepackage{xcolor}

\usepackage{cleveref}

\definecolor{cLs}{HTML}{FF9100}
\definecolor{cAm}{RGB}{255, 0.0, 255}
\definecolor{cCf}{RGB}{0, 255, 255}

\newcommand{\packageformat}[1]{\texttt{#1}}
\renewcommand{\emph}[1]{\textit{#1}}
\newcommand{\minmax}{\mathrm{minmax}}
\newcommand{\cLS}{\mathrm{LS}}        
\newcommand{\cCart}{\mathrm{C}}       
\newcommand{\cPol}{\mathrm{P}}        
\newcommand{\cLSP}{\mathrm{LS'}}      

\creflabelformat{equation}{#2\textup{#1}#3} 
\creflabelformat{figure}{#2#1#3}

\crefname{appendix}{}{}
\Crefname{appendix}{}{}
\crefname{algorithm}{Alg.}{Algorithms}
\Crefname{algorithm}{Algorithm}{Algorithms}
\crefname{equation}{Eq.}{Equations}
\Crefname{equation}{Equation}{Equations}
\crefname{figure}{Fig.}{Figures}
\Crefname{figure}{Figure}{Figures}
\crefname{listing}{Lst.}{Listings}
\Crefname{listing}{Listing}{Listings}
\crefname{section}{Sec.}{Sections}
\Crefname{section}{Section}{Sections}
\crefname{table}{Tab.}{Tables}
\Crefname{table}{Table}{Tables}

\crefformat{algorithm}{Alg.\,#2#1#3}
\Crefformat{algorithm}{Algorithm\,#2#1#3}
\crefformat{appendix}{#2#1#3}
\Crefformat{appendix}{#2#1#3}
\crefformat{equation}{Eq.\,#2#1#3}
\Crefformat{equation}{Equation\,#2#1#3}
\crefformat{figure}{Fig.\,#2#1#3}
\Crefformat{figure}{Figure\,#2#1#3}
\crefformat{listing}{Lst.\,#2#1#3}
\Crefformat{listing}{Listing\,#2#1#3}
\crefformat{section}{Sec.\,#2#1#3}
\Crefformat{section}{Section\,#2#1#3}
\crefformat{table}{Tab.\,#2#1#3}
\Crefformat{table}{Table\,#2#1#3}

\newlist{enumerateinline}{enumerate*}{1}
\setlist[enumerateinline]{label=(\alph*)}

\newacronym[
    description={
        Artificial Intelligence.\\
        Field of computer science focussing on building algorithms and systems that can perform actions and task that usually require human intelligence
    },
]
{ai}
{AI}
{artificial intelligence}

\newacronym[
    description={
        Neural Network.\\
        Deep AI model inspired by the human brain.
        Consists of nodes interconnected with weights that get adjusted during training to learn a mean best model for the data at hand 
    },
]
{ann}
{ANN}
{artificial neural network}

\newacronym[
    description={
        Autoencoder.\\
        Neural network to learn efficient encodings of datasets.
        Learns by mapping a series onto itself after passing it through a low-dimensional bottleneck layer
    },
]
{autoencoder}
{AE}
{Autoencoder}

\newacronym[
    description={
        Neural Network.\\
        Homepage: \url{https://brian2.readthedocs.io}.\\
        Simulation framework for spiking neural networks.
        Details in \cite{Stimberg2019_BRIAN2}.
    },
]
{brianii}
{Brian 2}
{Brian 2}
\glsunset{brianii}

\newacronym[
    description={
    },
    sort=dataanalysis,
]
{dataanalysis}
{data-analysis}
{data-analysis}
\glsunset{dataanalysis}

\newacronym[
    description={
    },
    sort=dataprocessing,
]
{dataprocessing}
{data-processing}
{data-processing}
\glsunset{dataprocessing}

\newacronym[
    description={
        Deep Learning.\\
        Artificial intelligence facilitated by deep neural networks (neural networks with 2 or more hidden layers).
    },
]
{dl}
{DL}
{deep learning}

\newacronym[
    description={
        Dark Energy Survey.\\
        Homepage: \url{https://www.darkenergysurvey.org/}.\\
        Modern astronomical survey monitoring the southern night-sky.
        Details in \citet{Bernstein2012_DES} and references therein
    },
    sort=des,
]
{des}
{DES}
{Dark Energy Survey}
\glsunset{des}

\newacronym[
    description={
        Dark Energy Spectroscopic Instrument.\\
        Homepage: \url{https://www.desi.lbl.gov//}.\\
        Instrument to measure the effect of dark energy on the expansion of the universe.
        Goal: obtain optical spectra for constructing a 3D map of the nearby universe.
        Details in \citet{Levi2019_DESI} and references therein
    },
    sort=desi,
]
{desi}
{DESI}
{Dark Energy Spectroscopic Instrument}
\glsunset{desi}

\newacronym[
    description={
        Extended LSST Astronomical Time-series Classification Challenge.\\
        Dataset for a community challenge to test brokers for LSST suitability
    },
    sort={ELAsTiCC},
]
{elasticc}
{ELAsTiCC}
{Extended LSST Astronomical Time-series Classification Challenge}
\glsunset{elasticc}

\newacronym[
    description={
        \textsc{Fink}.\\
        Homepage: \url{https://fink-broker.org}\\
        Community broker to process full stream of transient alerts from LSST
    },
    sort={fink},
]
{fink}
{\textsc{Fink}}
{\textsc{Fink}}
\glsunset{fink}

\newacronym[
    description={
        GitHub.\\
        Homepage: \url{https://github.com/}.\\
    },
]
{github}
{GitHub}
{GitHub}
\glsunset{github}

\newacronym[
    description={
        JavaScript.\\
        Programming language widely used in web development.
    },
]
{javascript}
{JavaScript}
{JavaScript}
\glsunset{javascript}

\newacronym[
    description={
        JavaScript Object Notation.\\
        File format based on java script syntax.
        Machine- as well as human-readable
    },
]
{json}
{JSON}
{Java Script Object Notation}
\glsunset{json}

\newacronym[
    description={
        Lightcurve.\\
        Brightness recordings of some astronomical object over time
    },
]
{lc}
{LC}
{lightcurve}

\newacronym[
    description={
        Rubin Legacy Survey of Space and Time.\\
        Homepage: \url{https://www.lsst.org/}.\\
        Modern astronomical survey monitoring the southern night-sky.
        Details in \citet{Ivezic2019_LSSTOverview}
    },
    sort=lsst,
]
{lsst}
{Rubin LSST}
{Vera C. Rubin Observatory Legacy Survey of Space and Time}
\glsunset{lsst}

\newacronym[
    description={
        Rubin Legacy Survey of Space and Time.\\
        Homepage: \url{https://github.com/TheRedElement/LStein}
    },
    sort=lstein,
]
{lstein}
{\textit{LStein}}
{\textit{Linking Series to envision information neatly}}
\glsunset{lstein}

\newacronym[
    description={
        MeerKAT.\\
        Radio telescope located in Meerkat National Park in South Africa.
        Formerly Karoo Array Telescope.
        Homepage: \url{https://www.sarao.ac.za/science/meerkat/about-meerkat/}.\\
        Details in \citet{Johnston2020_MeerKAT}
    },
    sort=meerkat,
]
{meerkat}
{MeerKAT}
{Karoo Array Telescope}
\glsunset{meerkat}

\newacronym[
    description={
        Machine Learning.\\
        A sub-field of AI focussed on algorithms and systems that adapt and learn based on data and patterns contained within
    },    
]
{ml}
{ML}
{machine learning}

\newacronym[
    description={
        Modified National Institute of Standards and Technology database.\\
        Database of handwritten digits.
    },    
]
{mnist}
{MNIST}
{modified National Institute of Standards and Technology database}

\newacronym[
    description={
        Portable Document Format.\\
    },
    sort=pdf,
]
{pdf}
{PDF}
{Portable Document Format}

\newacronym[
    description={
        \packageformat{matplotlib}.\\
        Homepage: \url{https://matplotlib.org/}.\\ 
        References: \citet{PY_Hunter2007_matplotlib}.\\
    },
    sort=matplotlib,
]
{pymatplotlib}
{\packageformat{matplotlib}}
{\packageformat{matplotlib}}
\glsunset{pymatplotlib}

\newacronym[
    description={
        \packageformat{sncosmo}.\\
        Homepage: \url{https://sncosmo.readthedocs.io/en/stable/}.\\ 
        References: \citet{PY_Barbary2025_sncosmo}.\\
    },
    sort=sncosmo,
]
{pysncosmo}
{\packageformat{sncosmo}}
{\packageformat{sncosmo}}
\glsunset{pysncosmo}

\newacronym[
    description={
        \packageformat{plotly}.\\
        Homepage: \url{https://plotly.com/}, \url{https://plotly.com/python/}.\\ 
        References: \citet{Plotly2015_Plotly}.\\
    },
    sort=plotly,
]
{plotly}
{\packageformat{Plotly}}
{\packageformat{Plotly}}
\glsunset{plotly}

\newacronym[
    description={
        Python.\\
        Homepage: \url{https://www.python.org/}.\\
        References: \citet{VanRossum2009_Python3}.\\
    },
]
{python}
{Python}
{Python}
\glsunset{python}

\newacronym[
    description={
        Read the Docs.\\
        Homepage: \url{https://about.readthedocs.com/}.\\
    },
    sort=readthedocs,
]
{readthedocs}
{Read the Docs}
{Read the Docs}
\glsunset{readthedocs}

\newacronym[
    description={
        Rubin Observatory.\\
        Homepage: \url{https://rubinobservatory.org/}.\\
        Telescope completing the LSST.
    },
    sort=rubin,
]
{rubin}
{Rubin}
{Rubin Observatory}
\glsunset{rubin}

\newacronym[
    description={
        Spectral Energy Distribution.\\
        Intensity of electromagnetic radiation of a source over different wavelengths.
    },
    \glslongpluralkey=spectral energy distribution,
    \glsshortpluralkey=SEDs,
]
{sed}
{SED}
{spectral energy distribution} 

\newacronym[
    description={
        Supernova.\\
        Explosive end of a stars life.
        The specific type depends on the evolutionary history and a stars core mass
    },
    \glslongpluralkey=supernovae,
    \glsshortpluralkey=SNe,
]
{sn}
{SN}
{supernova}

\newacronym[
    description={
        Spiking Neural Network.\\
        Biology inspired neural network extensively studie in computational neuroscience.
        Based on information relay via spike emission of neurons.
    },    
]
{snn}
{SNN}
{Spiking Neural Network}

\newacronym[
    description={
        Tidal Disruption Event.\\
        Transient event in which tidal forces of some astronomical object rip apart an orbiting object.
    },    
]
{tde}
{TDE}
{Tidal Disruption Event}

\newacronym[
    description={
        Zwicky Transient Facility.\\
        Homepage: \url{https://www.ztf.caltech.edu/}
    },
    sort=ztf,
]
{ztf}
{ZTF}
{Zwicky Transient Facility}
\glsunset{ztf}

\definecolor{cLinkText}{HTML}{1C8DB4}
\hypersetup{
	plainpages=false,			    
	colorlinks=true,			    
	citecolor=cLinkText,             
    linkcolor=cLinkText,             
    urlcolor=cLinkText,              
    pdfborder={0 0 0},			    
	breaklinks=true,			    
	bookmarksnumbered=true,		    
	bookmarksopen=true,			    
}

\definecolor{cCodeBg}{HTML}{D6D6D6}
\definecolor{cCodeConstant}{HTML}{000000}
\definecolor{cCodeComment}{HTML}{6A9955}
\definecolor{cCodeString}{HTML}{D16969}
\definecolor{cCodeNumeric}{HTML}{B5CEA8}
\definecolor{cCodeKeyword}{HTML}{569CD6}
\lstdefinestyle{vscodedarkstyle}{
	language=python,
    frame=Tb,
    basicstyle=\small\color{cCodeConstant},
    commentstyle=\color{cCodeComment},
    keywordstyle=\color{cCodeKeyword},
    numberstyle=\scriptsize\color{cCodeNumeric},
    stringstyle=\color{cCodeString},
    numbers=left,
    classoffset=1,
    breaklines=true,
    postbreak=\usebox\mypostbreak,
    upquote=true,
}
\newsavebox\mypostbreak
\savebox\mypostbreak{\raisebox{0ex}[0ex][0ex]{\ensuremath{\color{red}\hookrightarrow\space}}}
\lstdefinelanguage{JavaScript}{
    keywords={break, case, catch, continue, debugger, default, delete, do, else, finally,
        for, function, if, in, instanceof, new, return, switch, this, throw, try, typeof, var, let, const, while, with},
    sensitive=true,
    comment=[l]{//},
    morecomment=[s]{/*}{*/},
    morestring=[b]",
    morestring=[b]'
}

\renewcommand{\toprule}{
    \specialrule{\heavyrulewidth}{0pt}{\heavyrulewidth}
    \specialrule{\heavyrulewidth}{0pt}{\belowrulesep}
}

\begin{document}

\begin{frontmatter}

\title{LStein: A new approach to visualizing sparse 2.5-dimensional data.\tnoteref{fn-code}} 
\tnotetext[fn-code]{
    Source code is available on \gls{github}: \url{https://github.com/TheRedElement/LStein}.
}

\author[cas,ozgrav]{Lukas Steinwender\corref{fn-cor}} 
\ead{lsteinwender@swin.edu.au}
\ead[https://lukassteinwender.com]{https://lukassteinwender.com}
\cortext[fn-cor]{Correspondence to:
    Centre for Astrophysics \& Supercomputing, Swinburne University of Technology, John Street, Melbourne, 3122, Victoria, Australia.
    Tel.: +61\,3\,9214\,8444.
}
\author[cas,ozgrav]{Anais M\"{o}ller}
\ead{amoller@swin.edu.au}
\author[cas]{Christopher J.~Fluke}
\ead{cfluke@swin.edu.au}

\affiliation[cas]{organization={Centre for Astrophysics \& Supercomputing, Swinburne University of Technology},
            addressline={John Street},
            city={Melbourne},
            postcode={3122},
            state={Victoria},
            country={Australia}}
\affiliation[ozgrav]{organization={ARC Centre of Excellence for Gravitational Wave Discovery (OzGrav)},
            addressline={John Street},
            city={Hawthorn},
            postcode={3122},
            state={Victoria},
            country={Australia}}

\begin{abstract}
Visualization of high-dimensional data is crucial to retrieve all the knowledge that is contained within a dataset.
Effective and informative presentation of three-dimensional data via a two-dimensional medium is challenging, especially if the dataset more closely resembles a 2.5-dimensional (2.5D) entity due to sparse sampling.
We present \acrshort{lstein} (\acrlong{lstein}), a novel visualization approach implemented in \gls{python}, in an attempt to solve this challenge.
Inspired by the astrophysical application of displaying photometric timeseries in multiple passbands with minimal loss of information, we compare our method to traditional approaches.
While astronomy -- specifically multi-passband visualization for \acrlongpl{lc} obtained with the \acrlong{rubin} -- serves as the principal driver for the design, we demonstrate that \acrshort{lstein} can be used in any context with 2.5D datasets from radio astronomy to \acrlong{ml} hyperparameter search visualization.
\Gls{lstein} provides a complementary visualization to traditional techniques.
\Gls{lstein} can be installed from \gls{github} (\url{https://github.com/TheRedElement/LStein}).
\end{abstract}


\end{frontmatter}

\section{Introduction}
\label{sec-introduction}

Data-analysis and knowledge-discovery is intimately linked to visual inspection.
While 2-dimensional (2D) datasets are easily visualized using physical or digital 2D media without loss of information, this becomes more challenging for higher dimensions \citep{Fluke2009_Sharing3dGraphicsAstronomy}.
Consider 3-dimemsional (3D) datasets where projection effects distort the displayed information when presented in a 2D medium.
While interpolation can mitigate some of the lost information by linking close data points, for sparse 3D datasets this is a non-trivial process.

\begin{figure}[!h]
    \centering
    \includegraphics[width=\linewidth]{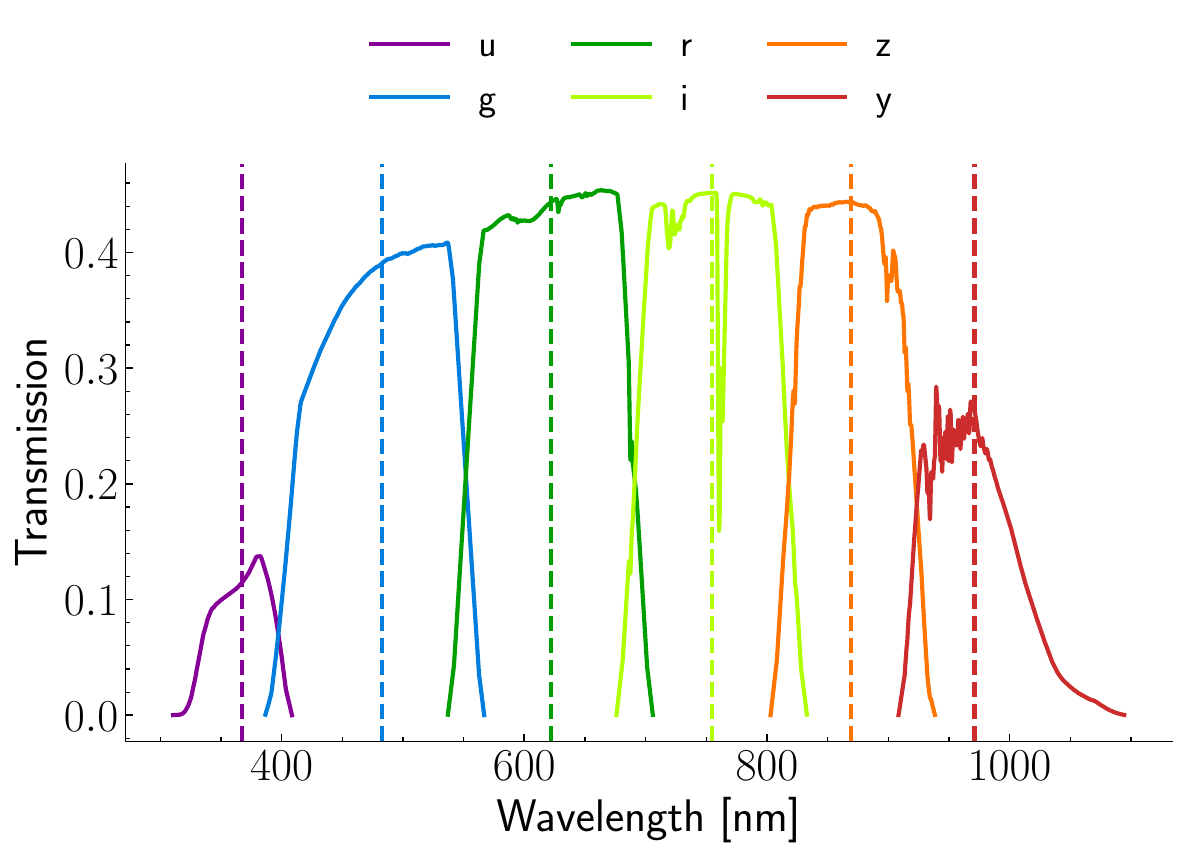}
    \caption{
        Transmission curves of the \gls{lsst} filter set.
        Note, that the passbands are not spaced uniformly in wavelength.
        Transmission data taken from \gls{pysncosmo} \citep{PY_Barbary2025_sncosmo}.
        We denote a passbands' average wavelength \citep[following][]{Koornneef1986_SyntheticPassbandPhotometry} with a dashed, vertical line.
    }
    \label{fig-filtercurve}
\end{figure}

\begin{figure}[!h]
    \centering
    \includegraphics[width=\linewidth]{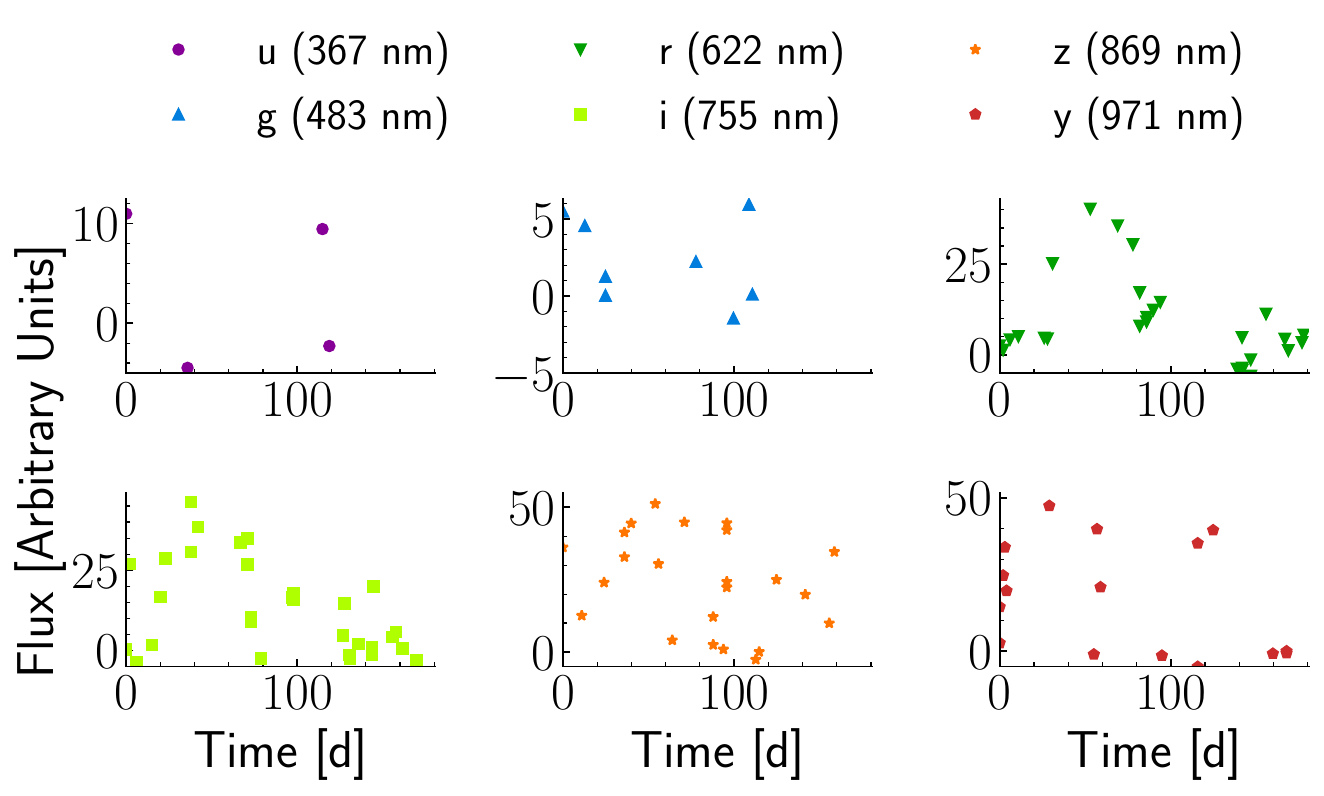}
    \caption{
        Simulated \gls{lc} of an \gls{elasticc} \gls{sn}.
        Each panel shows the variation in Flux (arbitrary units, without errors) over time for each of the passbands from \cref{fig-filtercurve}.
        Note the different scales of the $y$-axes, which are necessary to see the \gls{lc} morphologies.
        The only location where the passband-separation is recorded is in the legend.
    }
    \label{fig-multipanel}
\end{figure}

The example inspiring this work is time-dependent photometric observations in optical astronomy.
Here, astronomers observe the Universe in passbands (filters).
We show transmission curves from the \acrlong{lsst} \citep[\acrshort{lsst},][]{Ivezic2019_LSSTOverview,Hambleton2023_LSSTRoadmap} passbands in \cref{fig-filtercurve}.
Observing a signal over time will result in a unique curve for every passband (\cref{fig-multipanel}).
These brightness variations over time in a specific passband are referred to as \gls{lc}.
The curves shown in \cref{fig-multipanel} are from \acrshort{elasticc} \citep[\acrlong{elasticc},][]{Knop2023_ELAsTiCC}, a dataset of simulated \glspl{lc} for \gls{lsst}.

Considering \cref{fig-filtercurve} it is clear that the use of passbands only sparsely samples the wavelength dimension.
Furthermore, every measurement in a specific passband is a summary of the entire passbands transmission region, which is unique for each passband and can even overlap with other bands.
Surprisingly, this relation in wavelength-space is often overlooked, but essential to consider when making claims about the physics governing an object's behavior.
This is the case because the wavelength is intimately tied to physical phenomena, and the closeness of different passbands can give insight into the true landscape of wavelength-space.
Examples for the many science cases where the wavelength-dependent properties of photometric observations are crucial include supernova classification \citep[][]{Moeller2020_SuperNNova,Fraga2024_FinkElasticc,Moller2025_ActiveLearningSniaEarlyClassification}, asteroseismology \citep[][]{Balona1999_ModeIdentificationAsteroseismologyMulticolor,Belllinger2024_MulticolorAsteroseismology,Fritzewski2025_MulticolorAsteroseismology}, exoplanet analysis \citep[][]{PelaezTorres2024_MulticolorExoplanets}, and tidal disruption event discovery \citep{LlamasLanza2026_TdeEarlyClassification}.

\begin{figure*}[!t]
    \centering
    \includegraphics[width=0.54\linewidth]{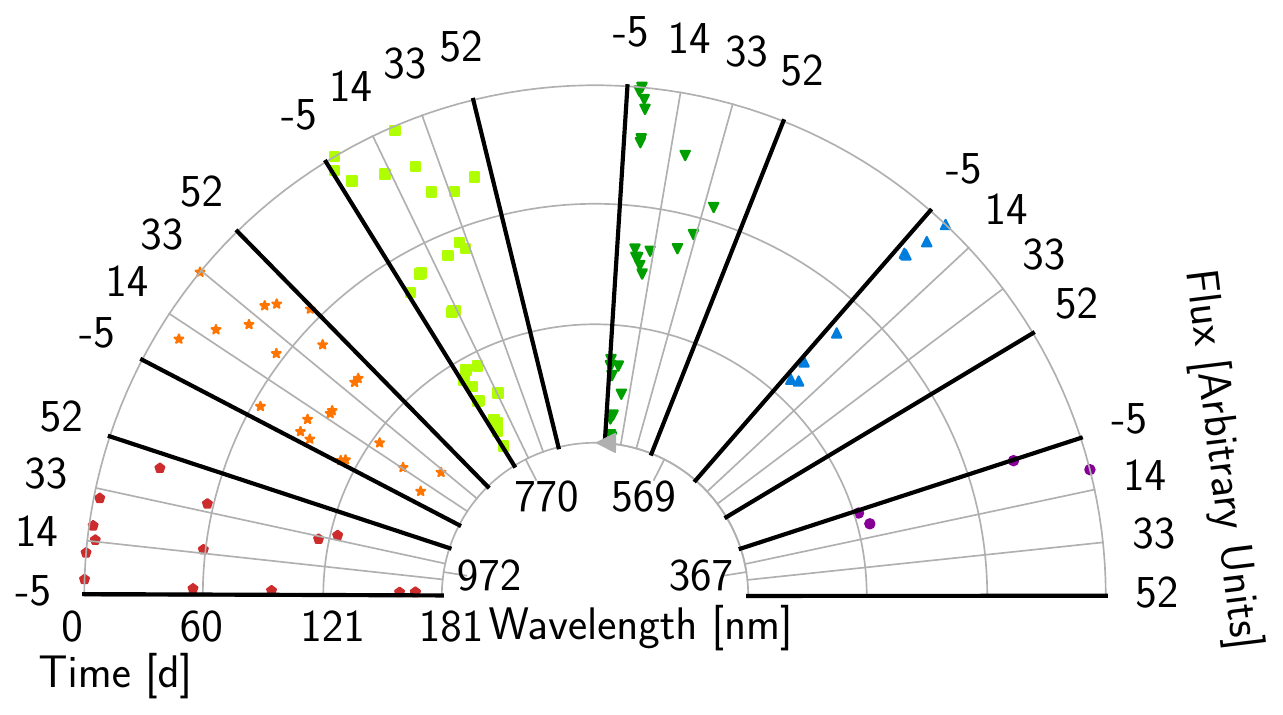}
    \includegraphics[width=0.44\linewidth]{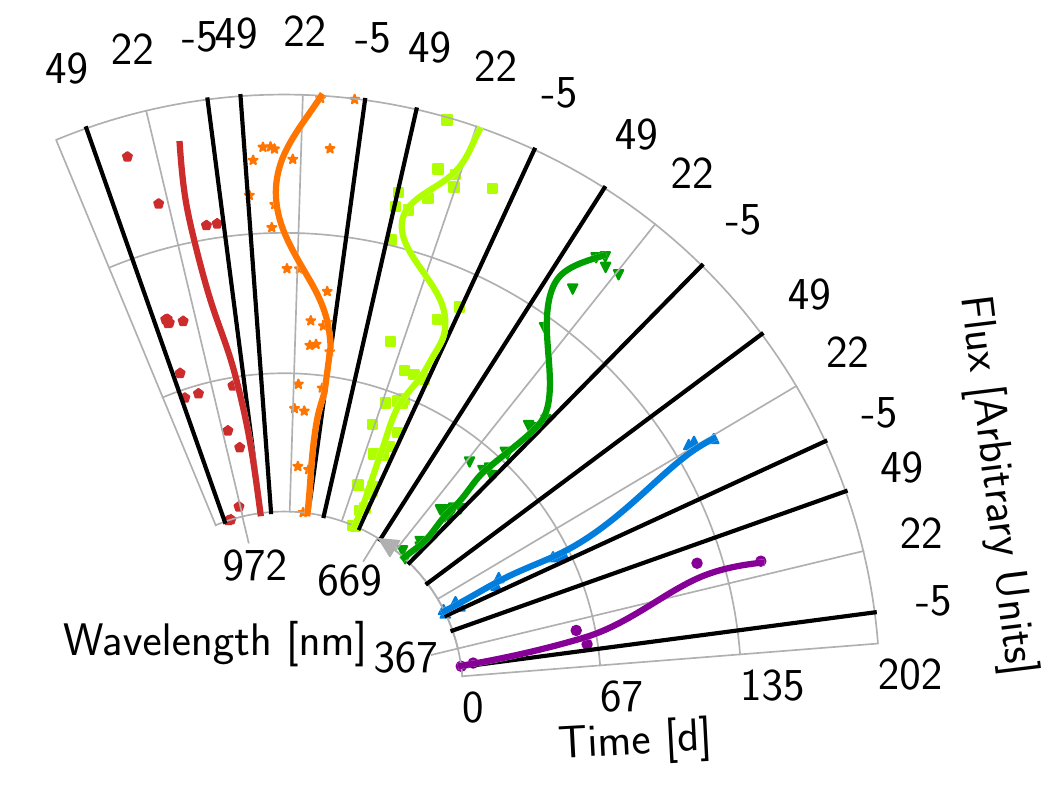}
    \caption{
        Example for \gls{lstein} plots.
        The left panel shows the application to a \gls{sn} \gls{lc}, the right panel to a \gls{tde}.
        Both \glspl{lc} are from \gls{elasticc} \citep[][]{Knop2023_ELAsTiCC}.
        The azimuthal axis (inner radius) encodes passband wavelength, the radial axis contains time, and the azimuthal sectors (outer axis) brightness.
        Sectors, denoted by thick black lines, are interpreted as individual panels.
        We also show representative curves in the right panel to demonstrate the capability of overplotting different series.
        To demonstrate customizability, we invert the time-, and flux-axis and adjust the space occupied by the plot for the right panel.
    }
    \label{fig-lsteinexamples}
\end{figure*}

To address the issues that arise when visualizing time-dependent photometric observations, we propose \gls{lstein}, a new, deterministic, way to visualize 2.5-dimensional (2.5D) data in two dimensions.
The 2D part is hereby the combination of time and brightness in our example.
Dimension 0.5 corresponds to wavelength.
We chose this terminology since the data has certainly more than 2 dimensions, but not quite 3 due to its sparseness in wavelength.
Our nomenclature shall not be confused with fractal dimensions \citep[i.e.,][]{Mandelbrot1967_FractalDimensions}.
An example application of \gls{lstein} can be found in \cref{fig-lsteinexamples}.

This paper is organized as follows:
In \cref{sec-traditionalmethods} we compare different traditional methods and highlight their shortcomings.
Context to the broader field of radial visualizations is provided in \cref{sec-relatedvisualizations}.
\Cref{sec-designprinciples} outlines the design principles that guided the implementation (\cref{sec-implementation}) of \gls{lstein}.
We demonstrate the broader adaptability and flexibility of \gls{lstein} by providing example applications for a range of different fields in \cref{sec-extensions}.
Finally, we summarize known issues with possible solution, and plans for future additions in \cref{sec-knownissuesandworkarounds,sec-furtherdevelopment}.
Combining \gls{lstein} with traditional visualization techniques can lead unique insights when visualizing 2.5D datasets.

\section{Traditional methods}
\label{sec-traditionalmethods}

We compare different traditional methods for 2.5D visualization by applying them to \gls{elasticc} (see \cref{sec-introduction}) \glspl{lc}.
Note, that the same \glspl{lc} are displayed for all comparisons.
Furthermore, keep in mind that colors are not necessary (different markers would suffice to differentiate between data-series) but were applied to make different series easily identifiable across figures.

\subsection{Multi-panel 2.5D visualization}
\label{sec-multipanel}
A natural approach to visualize a set of \glspl{lc} is to display each passband in a different panel (\cref{fig-multipanel}).
While the individual series are nicely distinguishable, the passband information is obscured due to the separation, which detaches the series from the wavelength information.
With this approach,  the wavelength spacing of the different passbands is not visualized.
Moreover, as the different data-series are detached from each-other, the viewer has to move from panel to panel to make comparisons of brightness and/or time values.

A further weakpoint of this approach is that certain panels might contain a lot of empty space, if some passbands show stronger recordings (higher brightness values) than others.
This, of course, only applies when all the axes are set to have the same limits (as has been done in \cref{fig-multipanel}).
If the axis limits are scaled based on the contained data, another issue occurs: the amplitudes are not comparable anymore.

\subsection{Single-panel 2.5D visualization}
\label{sec-singlepanel}
\begin{figure}[t]
    \centering
    \includegraphics[width=\linewidth]{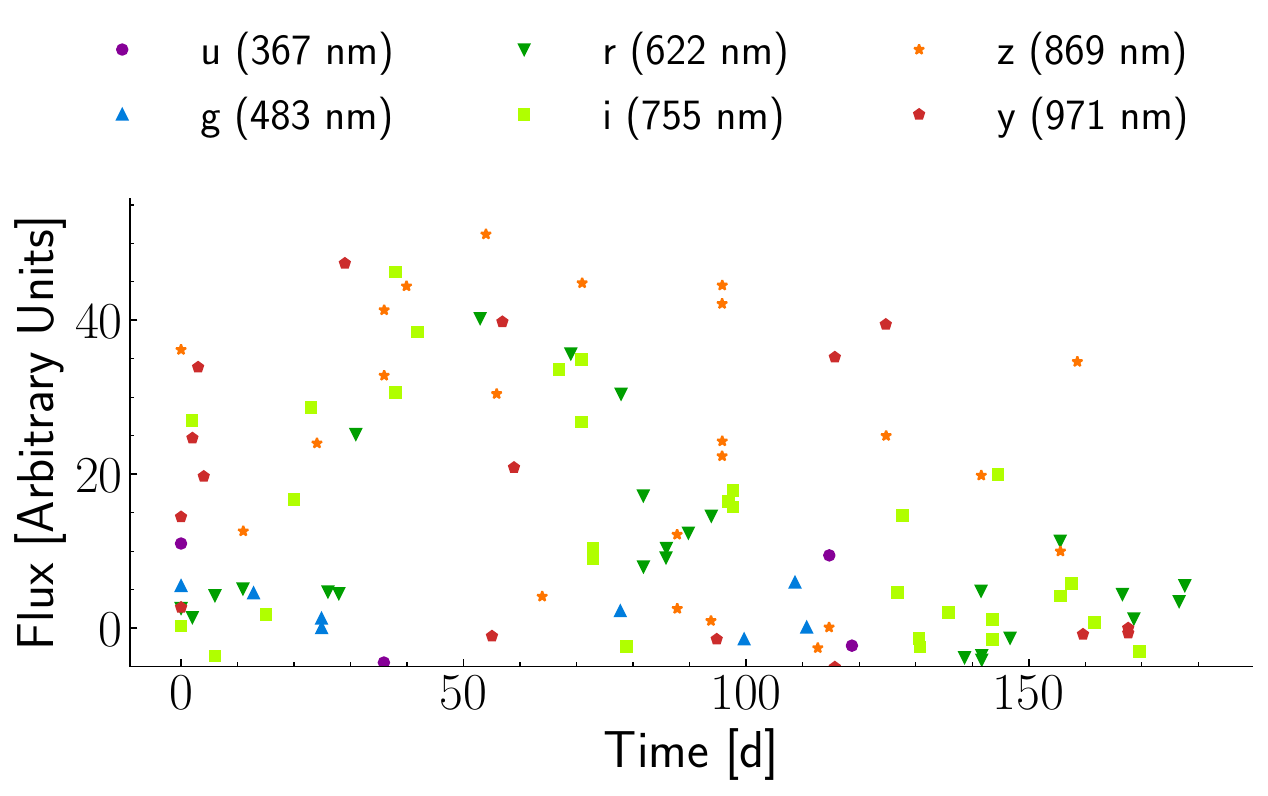}
    \caption{
        Example for the single panel approach.
        The same \gls{sn} and photometry as in \cref{fig-multipanel} is shown.
        Different colors denote different passbands.
    }
    \label{fig-onepanel}
\end{figure}

To address the series-detachment problem of the multi-panel approach (\cref{sec-multipanel}), the single-panel visualization displays all the information together.
While this works well with few passbands, it becomes more challenging as the number of passbands increases.
As can be observed in \cref{fig-onepanel}, even with only 6 passbands (as is the case for the \acrlong{rubin}) it is hard to disentangle the different data-series.
Comparing \cref{fig-onepanel} to \cref{fig-lsteinexamples} one can see that the different passband measurements as well as perception of maximum brightness are easier to disentangle.
The wavelength-dependence of the passbands is still not displayed, as the data points are just plotted on top of each other.

\subsection{Single panel with offset}
\label{sec-singlepaneloffset}

Another method of displaying \glspl{lc} is an extension to \cref{sec-singlepanel}, whereby all data points are plotted within a single panel, but an offset is applied to the respective series reflecting the corresponding passbands' wavelength.
This ensures that passband information is encoded in the vertical position of the \gls{lc}.
Drawbacks of this approach are that one has to choose between one of the following options:
\begin{enumerate}
    \item \emph{Display brightness values on the ordinate and hide passband information.}
        This has the downside that passband information has to be inferred from the relative offset between data-series reference points (e.g., median brightness, brightness-minimum).
        Therefore, the viewer is required to factor an additional piece of information into the analysis of the visualized data.
        Furthermore, depending on the actual wavelength values, it can easily happen that different series are overlapping, which might lead to confusion.
    \item \emph{Display passband information on the ordinate and hide brightness values.}
        Brightness, more specifically its change, is the main quantity of interest for many astrophysical analyses.
        Therefore, we argue that this choice is unsuitable by design.
    \item \emph{Display both quantities on the ordinate at the expense of a cluttered ordinate.}
        One of the main criteria of a good plot is readability.
        Too much text-information severely impacts readability of a visualization.
        Therefore, we assert that also this choice is unsuitable in most contexts.
\end{enumerate}
Furthermore, depending on the choice of the offset, it might be hard to get an intuition for the actual brightness values of each \gls{lc}.

Finally, comparing different objects can be very challenging because the passband-information is linked to the vertical offset/brightness information.
Therefore, a series with consistently huge brightness recordings will swallow the passband information, while series with tiny recordings get dominated by passband wavelength.

\begin{figure}[t]
    \centering
    \includegraphics[width=\linewidth]{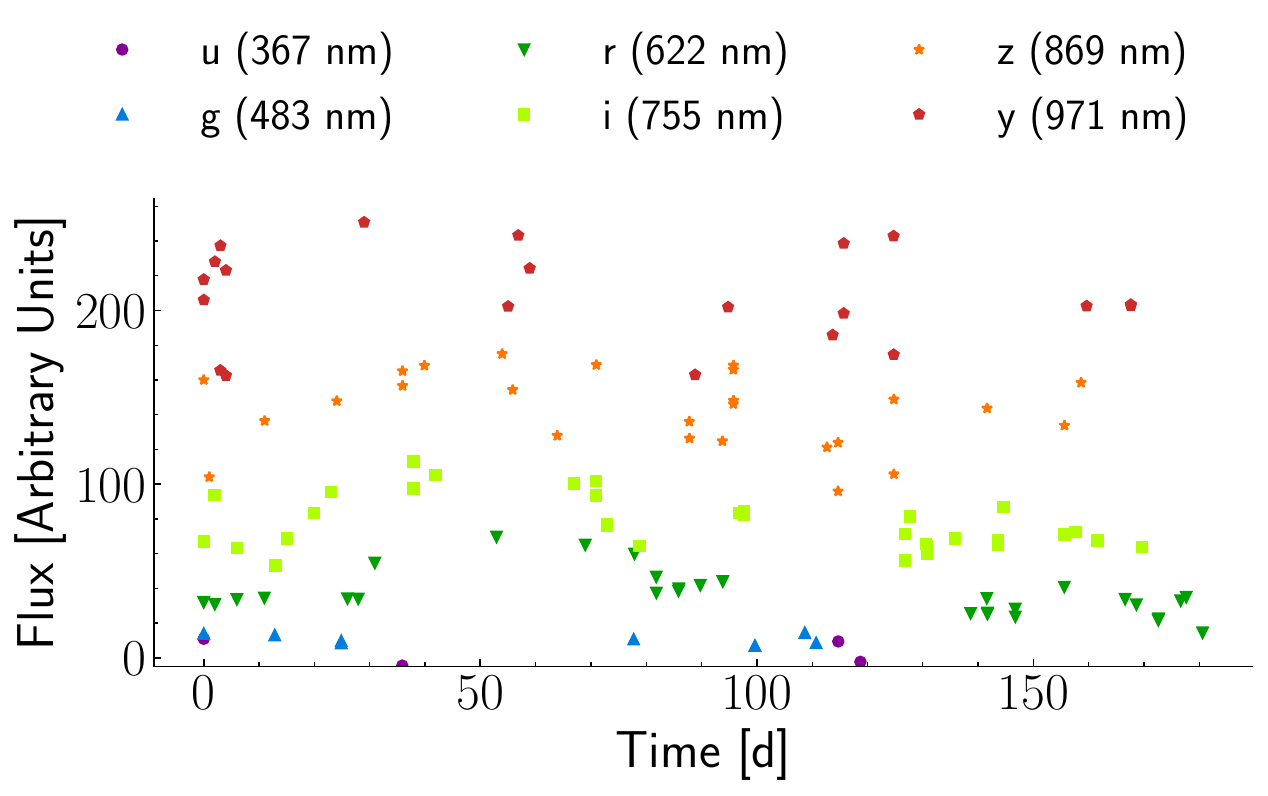}
    \caption{
        Example for displaying all dataseries in a single panel and encoding the wavelength information in the vertical offset.
        This particular example uses an offset of $\tfrac{\lambda}{100}$, where $\lambda$ is the passband wavelength.
        The offset is applied to the minimum of each dataseries.
        The same \gls{sn} and photometry as in \cref{fig-multipanel} is shown.
    }
    \label{fig-singlepaneloffset}
\end{figure}

\subsection{3D-plot}
\label{sec-meshsurface}

\begin{figure}[!t]
    \centering
    \includegraphics[width=1.0\linewidth]{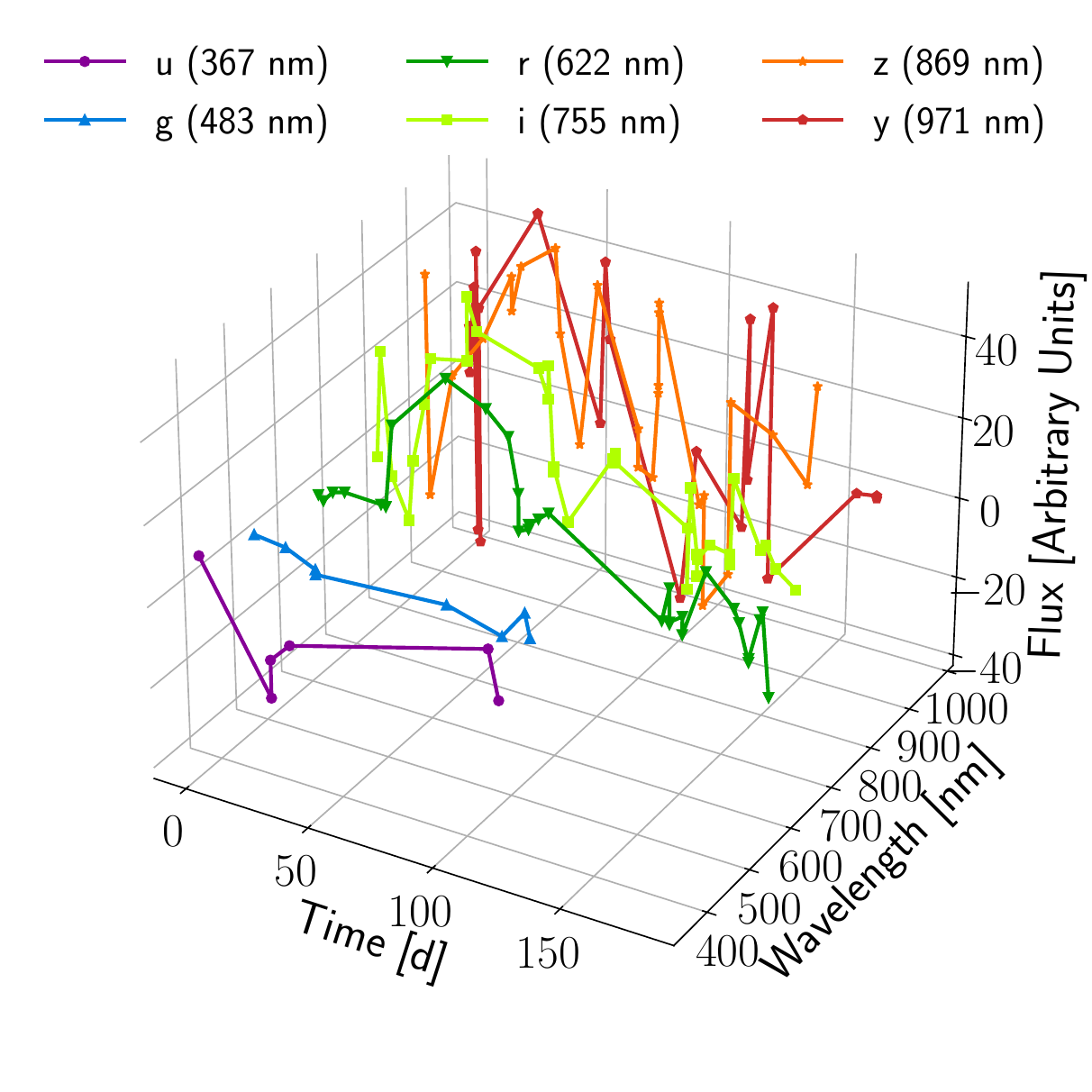}
    \caption{
        Example for a 3D visualization.
        We connect points with lines to make their location in 3D space more obvious.
        The same \gls{sn} and photometry as in \cref{fig-multipanel} is shown.
    }
    \label{fig-mesh}
\end{figure}

Transferring the problem into 3D space allows the inclusion of wavelength information in the display.
However, projection effects, plotting order, overlapping elements, and the fact that not all angles can be displayed hinder the easy interpretation when shared via a 2D medium (see \cref{fig-mesh} for an example).
Seeing as journals are the main medium to share research, \cite{Fluke2009_Sharing3dGraphicsAstronomy} attempted to alleviate these issues through the inclusion of interactive elements in \gls{pdf}-documents.
Another way to make 3D visualization more viable in the 2D publishing setting was explored by \cite{Russeil2024_Rainbow}, who attempted to add physical meaning to an interpolation and display the result as a smooth surface.

We note that inherently 3D displays such as virtual reality (VR) via head-mounted displays (HMDs), stereoscopic methods, and 3D printed output do not suffer from the drawbacks outlined for the projection of 3D space to 2D (\cref{fig-mesh}).
A discussion of these techniques is, however, out of the scope of this work because the main media for publishing data are in 2D.

\subsection{Radial plots}
\label{sec-relatedvisualizations}

\Gls{lstein} can be classified as a radial plotting techniques as demonstrated in \cref{fig-lsteinexamples}.
These are techniques that utilize circular or elliptical layouts to display data around some central point \citep{Burch2014_RadialDiagrams}.

Within the broader field of quantitative data visualization, there have been several recent studies investigating radial plots as an alternative to linear visualizations \citep[e.g.,][and references therein]{Waldner2020_RadialVsLinearCharts,Xie2024_RadialPlots}.
While \citet{Waldner2020_RadialVsLinearCharts} find that radial plots are generally less intuitive to interpret when working with 2D data, there are also arguments towards radial layouts like higher visual appeal and engagement.

Two examples from the astronomy community that exploit radial layouts include the starfish-diagram \citep[][used to position a object in the context of the whole population]{Konstantopoulos2015_StarfishDiagram} and TULIPS \citep[Tool for Understanding the Lives, Interiors, and Physics of Stars,][used for visualizations of stellar structure]{Laplace2022_Tulips}.
Both tools target different application cases than \gls{lstein}.
The ability to deal with a sparse dimension is something that sets \gls{lstein} apart from other radial layout methods.

\section{Design principles}
\label{sec-designprinciples}

The goal of \gls{lstein} is to create an \emph{easy to use}, \emph{customizable} framework for visualizing 2.5D data.
The fundamental idea is to project the 2.5D to 2D while preserving the (sparse) relationships that exist in the 0.5th dimension.
To illustrate the design principles for \gls{lstein}, we use the special case introduced in \cref{sec-introduction,sec-traditionalmethods}: multi-passband photometric \glspl{lc}.
To keep the terminology general we introduce a new coordinate system and relate our quantities of interest as follows:
\begin{align}
    \begin{split}
        \text{time}&\mapsto x^\cLS \\
        \text{brightness}&\mapsto y^\cLS \\
        \text{wavelength}&\mapsto \theta^\cLS
        \label{eq-coords}
    \end{split}
    .
\end{align}
We achieve the projection, by defining the following:
\begin{itemize}
    \item $x^\cLS$ in radial direction.
    \item $y^\cLS$ in azimuthal direction, but constrained to a sector of the entire $\theta^\cLS$ range.
    \item $\theta^\cLS$ in azimuthal direction.
\end{itemize}
Examples for \gls{lstein} plot-layouts are shown in \cref{fig-lsteinexamples}.

To set the $\theta^\cLS$-values in context, we make use of azimuthally-spaced panels.
Specifically, each $\theta^\cLS$-value is assigned to a panel that reflects its relation to all other $\theta^\cLS$-values.
Therefore, series that are closer in $\theta^\cLS$, are also physically plotted closer to each other.
The use of panels also reduces cluttering and makes sure that information such as series-colors or marker-shapes are only supplementary information, rather than serving as a separate data-axis.

We implement \gls{lstein} in \gls{python} \citep{VanRossum2009_Python3} to make sure it is easy to use and adopt.
Another motivating factor for this choice was that \gls{lsst} and other modern surveys predominantly use \gls{python} in their pipelines.

Given that \gls{pymatplotlib} \citep{PY_Hunter2007_matplotlib} is one of the most popular plotting libraries in \gls{python}, we opted for a logical flow similar to \gls{pymatplotlib}'s rendering order:
\begin{enumerateinline}
    \item creating a figure (called \texttt{LSteinCanvas});
    \item adding axes/panels (called \texttt{LSteinPanel});
    \item attaching artists to the axes\footnote{Artists are a concept from \gls{pymatplotlib}, which encompass elements such as lines or scatters.}.
\end{enumerateinline}
This makes sure that utilization is straightforward, highly customizable, and extendable.
A code example for the standard procedure is shown in \cref{lst-lsteinstandard}.
\begin{lstlisting}[
    float,    
    caption={
        Standard usage of \gls{lstein}.
        We omit arguments for readability and refer to the documentation for details on the object signature.
    },
    label={lst-lsteinstandard},
    style=vscodedarkstyle,
]
#import package
from lstein import lstein

#setup a new canvas
LSC = lstein.LSteinCanvas(...)

#add a panel
LSP = LSC.add_panel(theta=...)

#add artists (line, scatter) to panel
LSP.plot(xvals, yvals, ...)

#produce graphical output
fig = lstein.draw(LSC)
\end{lstlisting}

\subsection{Backends}
\label{sec-backends}
For the user's convenience, we include implementations for two backends in the package: \gls{pymatplotlib} and \gls{plotly} \citep{Plotly2025_Plotly}.
These can be invoked via \texttt{lstein.draw(LSC, backend="matplotlib")} and \texttt{lstein.draw(LSC, backend="plotly")}, respectively.

The core-implementation of \gls{lstein} is targeted towards using the \gls{pymatplotlib} backend because of \gls{pymatplotlib}'s popularity across the community.
Furthermore, \gls{pymatplotlib} is lightweight in terms of memory because it is generating static, publication-ready figures.
At the moment there is no interactive implementation of \gls{lstein} using the \gls{pymatplotlib} backend.

The \gls{plotly} backend has the advantage of producing interactive figures, which can also easily be incorporated into a web-page (\cref{sec-webintegration}).
We also added custom listeners in the \gls{plotly} backend implementation on top of \gls{plotly}'s default interactive elements.
These display contextual information about the data points through mouse-pointer hovering, providing direct access to data values.
If needed, further interactive elements can be included.
However, because \gls{pymatplotlib} is the core backend, the \gls{plotly} backend uses a translation of \gls{pymatplotlib} arguments to the \gls{plotly} framework.
For that reason, not all arguments will work as expected (\cref{sec-backendsubtleties}).
Furthermore, \gls{plotly} can require more memory and time for plotting because of its interactivity.

\Gls{lstein} is designed to be easily be adapted to arbitrary plotting frameworks beyond \gls{pymatplotlib} and \gls{plotly}.
This is achieved by using the \texttt{LSteinCanvas} instance as a container that has methods for computing the required axes and projections.
The most basic recipe to use \gls{lstein} with an arbitrary backend (\cref{lst-lsteinrecipe}) is shown in \cref{sec-additionalcodeexamples}.

\subsection{Customizations}
\label{sec-customizations}
Both \texttt{LSteinCanvas} and \texttt{LSteinPanel} can be customized to the user's liking by passing relevant keyword arguments.
We also provide convenience methods that will automatically create as many panels as needed (unique $\theta^\cLS$ present in \texttt{thetavals}).
This usage is outlined in \cref{lst-lsteinconvenience}, whereby \texttt{thetavals} is a 1D list or array, and \texttt{xvals} and \texttt{yvals} are lists of arrays (one array for each $\theta^\cLS$).
\begin{lstlisting}[
    float,    
    caption={
        Convenience usage of \gls{lstein}.
        We omit arguments for readability and refer to the documentation for details on the object signature.
    },
    label={lst-lsteinconvenience},
    style=vscodedarkstyle,
]
#import package
from lstein import lstein

#setup a new canvas
LSC = lstein.LSteinCanvas(...)

#add artists
LSC.plot(thetavals, xvals, yvals, ...)

#produce graphical output
fig = lstein.draw(LSC)
\end{lstlisting}

Note, that \texttt{lstein.draw(...)} is a convenience method to produce graphical output by invoking one of the provided backends (details in \cref{sec-backends}).
Additionally, we output the generated figure instance in the \texttt{lstein.draw()} method, enabling access to all low-level children of the backend in use.
Therefore, users are free to adjust the styling and layout of the plot as they please.

\subsection{Web integration}
\label{sec-webintegration}

\begin{figure*}[!t]
    \centering
    \includegraphics[width=\linewidth]{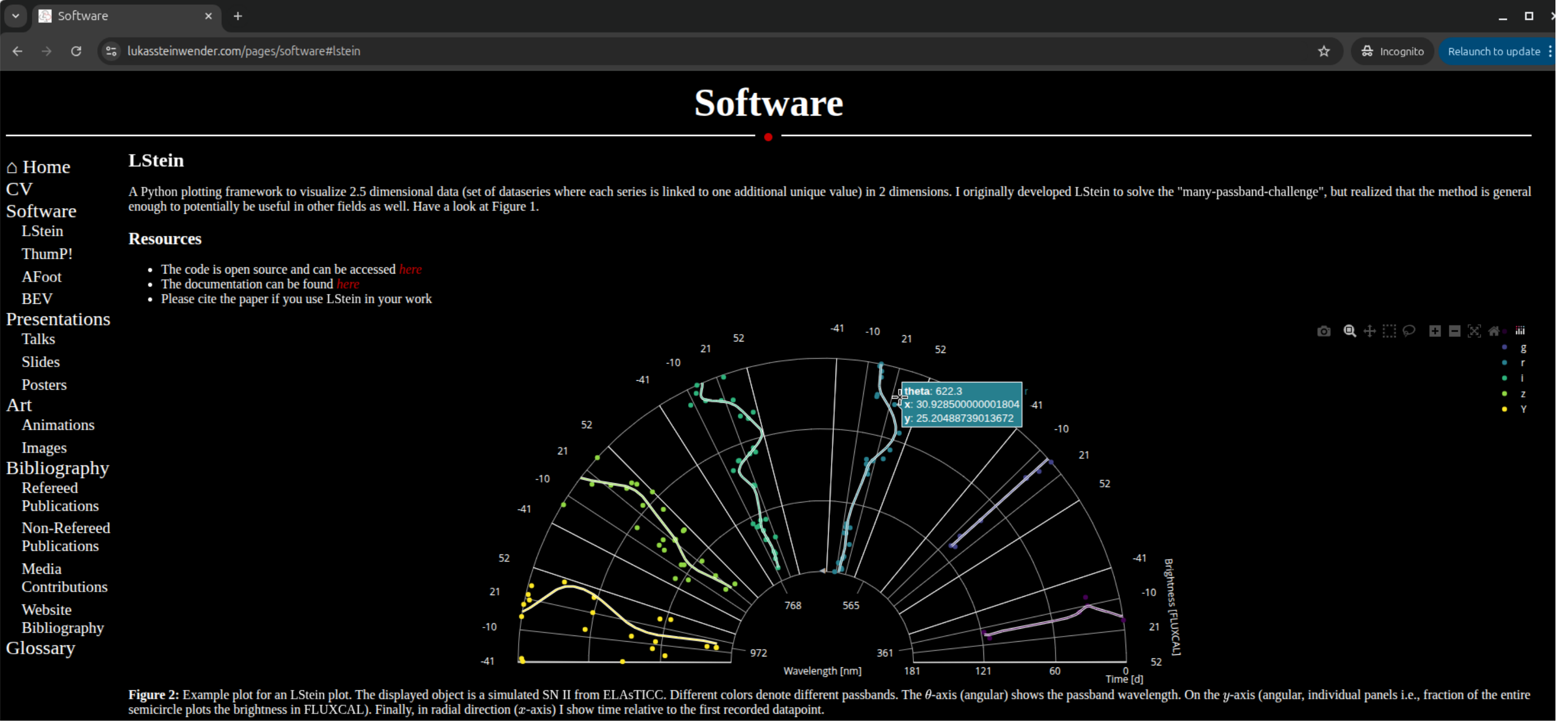}
    \caption{
        Screenshot of a \gls{lstein} plot integrated into a website.
        On the website, the plot is interactive as indicated by the text-box that appears on hovering any data-point.
    }
    \label{fig-webapplication}
\end{figure*}

Web integration is a powerful way to share visualizations and make them available for a broad audience.
\Gls{lstein} is especially well-suited for this task, as it allows all the relevant information to be displayed in a set amount of space.
The layout of the single panel with offsets approach (\cref{sec-singlepaneloffset}), for example, requires a tradeoff for many values of $\theta^\cLS$:
Less visible \gls{lc}-features, or an increase in the figures vertical size.
Due to the $\mathrm{2.5D} \mapsto \mathrm{2D}$ projection, \gls{lstein} even allows the artificial amplification of features, by placing them further away from the canvas-origin.

Web integration of \gls{lstein} also presents an easy way to share interactive versions of a created plot.
Through the implementation of the \gls{plotly} backend, we allow an easy way for users to create interactive plots.
These can be saved in a way that allows integration into a website\footnote{An example for a \gls{lstein} plot incorporated in a web-page can be found at \href{https://lukassteinwender.com/pages/software\#lstein}{https://lukassteinwender.com/pages/software\#lstein}} (\cref{fig-webapplication}).

Creating an interactive web element is as easy as
\begin{enumerateinline}
    \item using the \texttt{write\_json()} functions from the \texttt{plotly.io} module, which generates a \acrshort{json}-file (\acrlong{json});
    \item loading the saved \gls{json}-file in \gls{javascript};
    \item using \gls{plotly} within \gls{javascript} to generate the plot (code in \cref{lst-plotfromjson}).
\end{enumerateinline}

We are currently collaborating with the \acrlong{fink} broker \citep[not an acronym,][]{Moller2021_FINK} to integrate \gls{lstein} into their \gls{fink}-portal\footnote{\href{https://ztf.fink-portal.org}{ztf.fink-portal.org}, \href{https://lsst.fink-portal.org}{lsst.fink-portal.org}}.
\Gls{fink} is one of the \gls{lsst} community brokers that receives the time-domain data-stream, processes it, and supplies enriched and filtered data publicly to the community.
The inclusion of \gls{lstein} will allow users of \gls{fink} to get a different, supplementary view on data generated by \gls{lsst}.

\subsection{Data comparisons}
\label{sec-datacomparisons}
Finally, we aimed to enable an easy comparison between different data-series/instances of \gls{lstein}.
This is achieved in several ways.
First, the high level of customizability, for example, defining the displayed range of the $\theta^\cLS$-axis in combination with the capability of flipping different axes allows for a representation that can be adjusted to fit any specific dataset.
Additionally, the definition of constant ticks on each axis allows comparing different panels with the same layout by flipping back-and-forth between different plots.

\section{Implementation}
\label{sec-implementation}

\begin{figure}[!t]
    \centering
    \includegraphics[width=1.0\linewidth]{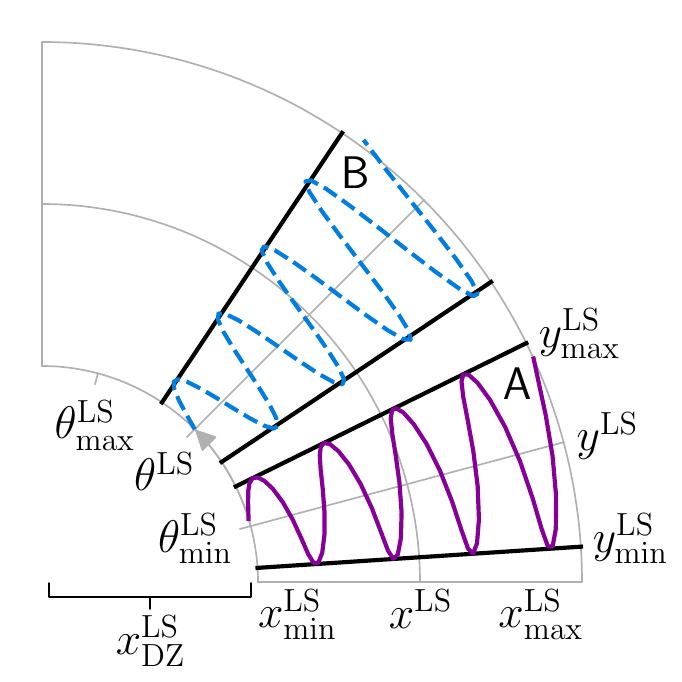}
    \caption{
        Definitions of the \gls{lstein} coordinate system.
        Gray arcs represent $x$-ticks, with $x^\cLS$ encoding different tick-values (corresponding to $x^\cCart$).
        Gray ticks on the innermost arc are $\theta$-ticks, the arrow on the innermost arc denotes the direction of $\theta$-ticks.
        The $\theta$-ticks don't necessarily have to align with displayed panels.
        The combination of $x$-ticks and $\theta$-ticks is referred to as ``fundament-grid'' of the \texttt{LSteinCanvas}.
        We denote panel-bounds with thick black lines (rays) connecting the innermost and outermost $x$-ticks.
        A \texttt{LSteinPanel} is then the annular sector enclosed within these panel-bounds.
        Gray rays within a \texttt{LSteinPanel} represent $y$-ticks, where $y^\cLS$ encodes the original $y^\cCart$.
        The empty space around the origin of the \texttt{LSteinCanvas} is referred to as $x^\cLS_\mathrm{DZ}$ ($x^\cLS$-deadzone) and reduces cluttering.
        All variables displayed show the actual physical information (\cref{sec-addingcontext}) but are plotted using $\cLSP$ coordinates (\cref{tab-projectionvariables,sec-projecttheta,sec-projecty}).
        The plotted data in panels A and B show results of projections using \texttt{y\_projection\_method="theta"} (\cref{sec-projecttheta}) and \texttt{y\_projection\_method="y"} (\cref{sec-projecty}), respectively.
    }
    \label{fig-lsteinvars}
\end{figure}

We follow the Pythonic object-oriented approach to ensure readable code and easy debugging, enable testing, and facilitate future extension of the package.
The two main objects that compose \gls{lstein} are \texttt{LSteinCanvas} and \texttt{LSteinPanel}.

The functionality of \texttt{LSteinCanvas} is to draw the annular sectors which define the $x^\cLS$-axis (radial arcs denote $x$-ticks) and ticks for the $\theta^\cLS$-axis (azimuthal direction).
The ticks simultaneously define the bounds for the respective axes.
\Cref{fig-lsteinvars} is a summary of the main elements.

\texttt{LSteinPanel}, on the other hand, deals with populating \texttt{LSteinCanvas} with panels, plotting their $y$-ticks and, most importantly, projecting any data-series into the \gls{lstein} coordinate system (\cref{fig-lsteinvars}).
Also here the $y$-ticks simultaneously define the axis bounds.

\subsection{Data-series projection}
\label{sec-dataseriesprojection}

We provide two methods for projecting data into the \gls{lstein} coordinate system: \texttt{theta} that completes the main projection step in polar coordinates, and \texttt{y} that operates in Cartesian coordinates for the main projection step.
In cases where it is crucial that the data is contained in the panel \texttt{theta} is preferred, if minimizing distortions is the main concern we recommend using \texttt{y}.
The basic idea of both is the same, however, the sequence of operations and some subtleties give different methods the edge in different use cases.
The methods used to project a single point into the \gls{lstein} coordinate system are described in \cref{sec-commonalities,sec-projecttheta,sec-projecty,sec-addingcontext}.
We summarize the variables used in the projections in \cref{tab-projectionvariables,fig-lsteinvars}.
Visualizations to aid the understanding of the different steps can be found in \cref{fig-projectionprep,fig-projectiontheta,fig-projectiony}.
Parameters superscribed with $\cCart$ denote the original (Cartesian) reference frame of the data.
We superscribe parameters used to actually plot the data points with $\cLS'$ and the general \gls{lstein} reference-frame with $\cLS$.
The latter includes also the physical relations present in the original reference frame.
\begin{table}[!t]
    \centering
    \caption{
        Variables used when projecting data points into the \gls{lstein}-context.
        The first column denotes the variable's symbol, the second column provides a brief description.
        We denote variables in the Cartesian context with a superscript $\cCart$ and variables within the \gls{lstein}-projection with superscript $\cLSP$.
        All the $\cLSP$ variables have counterparts that encode the actual physical meaning on the final plot ($\cLS$, as shown in text).
        These $\cLS$ variables can be thought of as a rescaled version of $\cLSP$ to match the input ranges in the Cartesian context.
    }
    \label{tab-projectionvariables}
    \begin{tabular}{p{0.3\linewidth}p{0.6\linewidth}}
        \toprule
        Variable & Description \\
        \midrule
        \multicolumn{2}{c}{Cartesian} \\
        $x^\cCart \in \mathbb{R}$                   & $x^\cCart$-value of a single observation in Cartesian space. \\
        $y^\cCart \in \mathbb{R}$                   & $y^\cCart$-value of a single observation in Cartesian space. \\
        $\theta^\cCart \in \mathbb{R}$              & 0.5-th dimension. \\
        $x^\cCart_{\min} \in \mathbb{R}$            & Minimum of $x^\cCart$. Lower limit of values in $x^\cCart$. \\
        $x^\cCart_{\max} \in \mathbb{R}$            & Maximum of $x^\cCart$. Upper limit of values in $x^\cCart$. \\
        $y^\cCart_{\min} \in \mathbb{R}$            & Minimum of $y^\cCart$. Lower limit of values in $y^\cCart$. \\
        $y^\cCart_{\max} \in \mathbb{R}$            & Maximum of $y^\cCart$. Upper limit of values in $y^\cCart$. \\
        $\theta^\cCart_{\min} \in \mathbb{R}$       & Minimum of $\theta^\cCart$. Lower limit of values in $\theta^\cCart$. \\
        $\theta^\cCart_{\max} \in \mathbb{R}$       & Maximum of $\theta^\cCart$. Upper limit of values in $\theta^\cCart$. \\
        \midrule
        \multicolumn{2}{c}{\gls{lstein}} \\
        $x^\cLSP \in [-1, 1]$                       & Cartesian $x$-coordinate that gets plotted in the \gls{lstein} panel. \\
        $y^\cLSP \in [-1, 1]$                       & Cartesian $y$-coordinate that gets plotted in the \gls{lstein} panel. \\
        $\theta^\cLSP \in [0,2\pi)$                 & Polar $\theta$-coordinate used to determine the panel offset. \\
        $x^\cLSP_\mathrm{DZ} \in [0,1]$             & fraction of $\vert x^\cLSP\vert$ that is not occupied by any data point. \\
        $\theta^\cLSP_{\min} \in [0,2\pi)$          & Minimum of $\theta^\cLSP$. Position of panel hosting $\theta^\cCart_{\min}$. \\
        $\theta^\cLSP_{\max} \in [0,2\pi)$          & Maximum of $\theta^\cLSP$. Position of panel hosting $\theta^\cCart_{\max}$. \\
        $\Delta\theta^\cLSP \in [0,2\pi)$           & Width of a single panel in the \gls{lstein} plot. \\
        \bottomrule
    \end{tabular}
\end{table}

\subsubsection{Commonalities across projection methods}
\label{sec-commonalities}
\begin{figure*}[!t]
    \centering
    \includegraphics[width=1.0\linewidth]{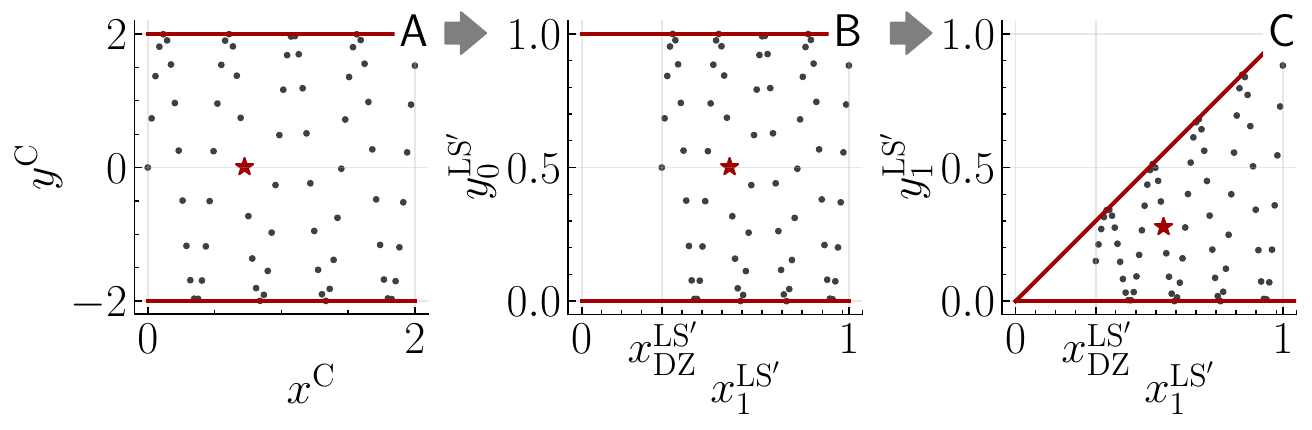}
    \caption{
        Visualization of the transformations applied before all other specific projection methods (\cref{fig-projectiontheta,fig-projectiony}).
        Red, solid lines indicate panel-bounds, gray arrows denote a transformation.
        The $y$-labels refer to the panels' $y$-axis (enclosed by red, solid lines).
        The red star is a randomly chosen point tracking the data-flow across figures throughout the transformation.
        Panel A contains the original data from Cartesian space with example data ranges of $x^\cCart \in [0,2]$ and $y^\cCart \in [-2, 2]$.
        Panel B applies $x^\cLSP_\mathrm{DZ}$ and projects to the $\ell_\infty$ unit ball (\cref{eq-xproj1,eq-yproj0}).
        Panel C makes sure that the projected $y$-values will fit into the \gls{lstein}-panel (\cref{eq-yproj1}).
    }
    \label{fig-projectionprep}
\end{figure*}

The following steps precede all specific projection methods (\cref{sec-projecttheta,sec-projecty}).
They can be thought of as preprocessing to make the projection behave as intended.
As such, \cref{fig-projectionprep} has to be considered as a precursor to \cref{fig-projectiontheta,fig-projectiony}, when interpreting.

\paragraph{Min-max-scaling}
Arguably the core-piece of \gls{lstein} and used in several steps to achieve the projection.
This operation rescales some value $z$ into a target-interval $[z'_{\min}, z'_{\max}]$ given the original interval it came from $[z_{\min},z_{\max}]$:
\begin{align}
    \begin{split}
        z'
            &= \minmax(z,
                z'_{\min}, z'_{\max},
                z_{\min}, z_{\max}
            ) \\
            &= \frac{z - z_{\min}}{z_{\max} - z_{\min}} \cdot (z'_{\max} - z'_{\min}) + z'_{\min}
    \end{split}
    .
    \label{eq-minmaxscale}
\end{align}

\paragraph{Panelbounds}
The entirety of \gls{lstein} operates inside the closed $\ell_\infty$ unit ball ($\{x\in \mathbb{R}^2: \Vert x\Vert_\infty \le 1\}$) to ensure well-behaving of trigonometric functions such as $\arctan$ and $\tan$, and avoid numerical instabilities in fractions.
To achieve this, one has to scale $x^\cCart$ and $y^\cCart$ accordingly.
Scaling $x^\cCart$ is achieved by applying \cref{eq-minmaxscale}:
\begin{align}
    x^\cLSP_1
        &= \minmax(x^\cCart,x^\cLSP_\mathrm{DZ},1,x^\cCart_{\min},x^\cCart_{\max})
    .
    \label{eq-xproj1}
\end{align}
We use $x^\cLSP_\mathrm{DZ}$ as the lower bound to make the plot less cluttered, which results in a region of empty space around the origin of \gls{lstein} ($x^\cLSP$-deadzone, see panel B in \cref{fig-projectionprep}).
The upper bound is a result of operating in the $\ell_\infty$ unit ball.
As for the adjustment of $y^\cCart$, we need to consider that \gls{lstein}-panels are sectors of an annulus (see \cref{fig-lsteinexamples,fig-lsteinvars} for examples).
Therefore, the range occupied by $y$-values at any given $x$-value has to scale linearly with the distance from the origin.
This is achieved in two steps.
First, $y^\cCart$ is rescaled to the unit-range (\cref{fig-projectionprep}, panel B):
\begin{align}
    y^\cLSP_0
        &= \minmax(y^\cCart,0,1,y^\cCart_{\min},y^\cCart_{\max})
    .
    \label{eq-yproj0}
\end{align}
Consecutively $x^\cLSP_1$ (\cref{eq-xproj1}) is applied to \cref{eq-yproj0} as a scaling factor (\cref{fig-projectionprep}, panel C):
\begin{align}
    y^\cLSP_1
        &= y^\cLSP_0 \cdot x^\cLSP_1
    .
    \label{eq-yproj1}
\end{align}

\paragraph{Panel-position computation}
To compute the correct placement of a panel, one converts the input $\theta^\cCart$ to an angle ($\theta^\cLSP$) that reflects the input range.
This is once more a matter of applying \cref{eq-minmaxscale}:
\begin{align}
    \theta^\cLSP
        &= \minmax(\theta^\cCart,
            \theta^\cLSP_{\min}, \theta^\cLSP_{\max},
            \theta^\cCart_{\min}, \theta^\cCart_{\max}
        )
    .
    \label{eq-thetaoffset}
\end{align}

\subsubsection{Projection method: \texttt{theta}}
\label{sec-projecttheta}
\begin{figure*}[!t]
    \centering
    \includegraphics[width=1.0\linewidth]{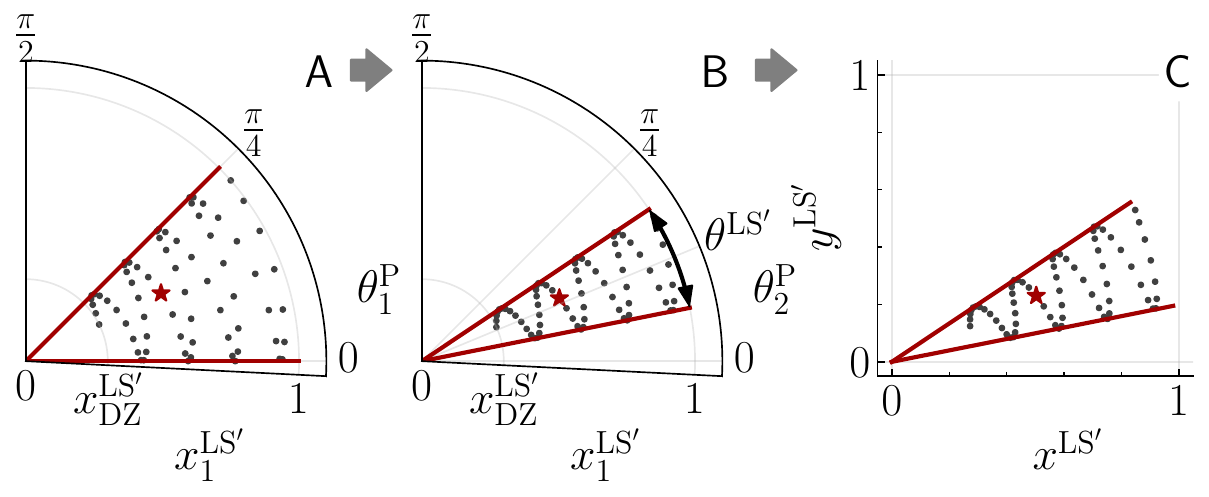}
    \caption{
        Steps for the projection into \gls{lstein} frame of reference using \texttt{y\_projection\_method="theta"} as described in \cref{sec-projecttheta}.
        Red, solid lines indicate panel-bounds, gray arrows denote a transformation.
        We represent $\Delta \theta^\cLSP$ (\cref{tab-projectionvariables}) with the black double-headed arrow.
        For panels A and B the $\theta$-labels refer to the panels' $\theta$-axis (enclosed by red, solid lines).
        The red star is the same randomly chosen point as in \cref{fig-projectionprep}, tracking the data-flow across figures throughout the transformation.
        Panel A is a transformation to polar coordinates.
        Panel B applies \cref{eq-minmaxscale} to place the data to the correct position (\cref{eq-thetapol2}). The panel is placed at $\theta^\cLSP$ (\cref{eq-thetaoffset}).
        Panel C shows the conversion back to Cartesian space for plotting as required by \gls{pymatplotlib} and \gls{plotly} (\cref{eq-projthetaresult}).
        All panels use $x^\cLSP_1$ as the value for $r$ when converting between Cartesian and polar coordinates to reduce projection effects in the $x$-direction.
    }
    \label{fig-projectiontheta}
\end{figure*}

This projection method basically guarantees that the plotted data will be contained in the corresponding \texttt{LSteinPanel}, which is especially useful for space-critical applications (e.g., web-displays, \cref{sec-webintegration}).
The tradeoff is that the projection effects are stronger compared to using \texttt{y} (\cref{sec-projecty}) as projection method.

For this projection method we start by transforming \cref{eq-xproj1,eq-yproj1} into polar coordinates, as this is the frame of reference for placing the data points into the correct panels (\cref{fig-projectiontheta}, panel A):
\begin{align}
    \theta^\cPol_1
        = \arctan\left(\frac{y^\cLSP_1}{x^\cLSP_1}\right)
    ,
    \label{eq-thetapol}
\end{align}
where the superscript $\cPol$ denotes a value in polar coordinates.
Rescaling the result from \cref{eq-thetapol} to the correct panel-bounds using \cref{eq-minmaxscale} places the data points at the desired position (panel B in \cref{fig-projectiontheta}):
\begin{align}
    \begin{split}
    \theta^\cPol_2
        = &\minmax(\theta^\cPol_1,
            \theta^\cLSP-\frac{\Delta\theta^\cLSP}{2}, \theta^\cLSP+\frac{\Delta\theta^\cLSP}{2},
            0, \frac{\pi}{4}
        )
    \label{eq-thetapol2}
    \end{split}
    ,
\end{align}
where $0$ and $\tfrac{\pi}{4}$ originate from choosing to operate in the closed $\ell_\infty$ unit ball ($\arctan\left(\tfrac{0}{1}\right)$ and $\arctan\left(\tfrac{1}{1}\right)$, respectively).
Depending on the implementation of $\arctan$, an additional offset of $\pi$ might be necessary.
Finally, we convert back to Cartesian coordinates for plotting (\cref{fig-projectiontheta}, panel C):
\begin{align}
    \begin{split}
        x^\cLSP &= x^\cLSP_1, \\
        y^\cLSP &= x^\cLSP \cdot \sin(\theta^\cPol_2)
        \label{eq-projthetaresult}
    \end{split}
    .
\end{align}
Take note that we do not use the polar coordinates' radius but rather $x^\cLSP$ for the back-conversion.
This is done to minimize projection effects in $x$-direction, since $y$ shall be encoded almost entirely in the angular part.

\subsubsection{Projection method: \texttt{y}}
\label{sec-projecty}

\begin{figure*}[!t]
    \centering
    \includegraphics[width=1.0\linewidth]{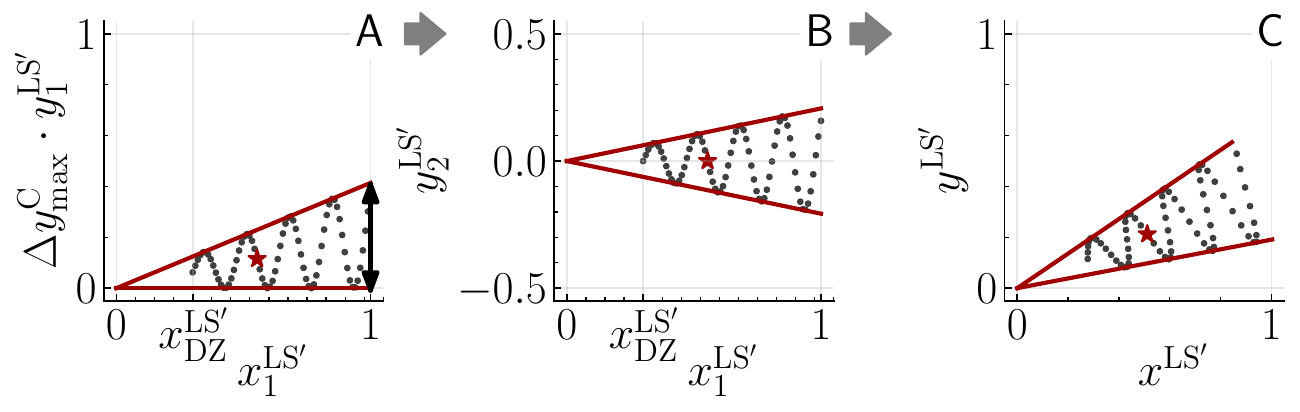}
    \caption{
        Steps of the projection using \texttt{y\_projection\_method="y"}.
        Red, solid lines indicate panel-bounds, gray arrows denote a transformation.
        We represent $\Delta y_{\max}^\cCart$ (\cref{eq-panelcartub}) with the black double-headed arrow.
        The red star is the same randomly chosen point as in \cref{fig-projectionprep}, tracking the data-flow across figures throughout the transformation.
        For panels A and B the $y$-labels refer to the panels' $y$-axis (enclosed by red, solid lines).
        Panel A projects the data to the correct angular width in Cartesian units (first term in \cref{eq-yproj2}).
        Panel B centers the projected sector, by applying the second term of \cref{eq-yproj2} to the result from panel A.
        Panel C places the data to the correct position by applying a rotation (\cref{eq-projypolar,eq-projyresult}).
    }
    \label{fig-projectiony}
\end{figure*}

Contrary to \cref{sec-projecttheta} (using \texttt{theta} as the projection method) one accepts that some points might overflow the panel for the advantage of minimzing distortion of the data.
We recommend using this method when the ranges of the data-series' $x$- and $y$-values are on completely different orders of magnitude, as projecting using \texttt{y} circumvents the use of $\arctan$ with a fraction of $x$ and $y$.

In this context we aim to make all our projections (specifically the projection into the correct panel bounds) in carthesian space and only use polar coordinates for the panel placement.
To do so, we project $y^\cLSP_1$ from \cref{eq-yproj1} into the correct angular width ($\Delta\theta^\cLSP$).
We start by computing the width of the annulus sector at any given $x^\cLSP_1$ in Cartesian coordinates ($\Delta y^\cCart$):
\begin{align}
    \Delta y^\cCart
        &= x^\cLSP_1 \cdot \tan(\Delta\theta^\cLSP)
    .
    \label{eq-panelcartsize}
\end{align}

Next, we determine the actual upper bound of the panel to ensure the data points get scaled correctly, even if they do not fill the entire axis-range:
\begin{align}
    \Delta y^\cCart_{\max}
        &= \max\left(\max(\Delta y^\cCart), 1 \cdot \tan(\Delta\theta^\cLSP)\right)
    .
    \label{eq-panelcartub}
\end{align}
The second argument in the $\max$ is hereby simply the upper bound of the panel based on the provided axis limits.
We can omit the contribution of the radius (set it to $1$) because we operate in the $\ell_\infty$ unit ball.

Consecutively, we combine \cref{eq-panelcartsize,eq-panelcartub,eq-yproj1} to get our data points projected into the \gls{lstein}-panel:
\begin{align}
    y^\cLSP_2
        &= \Delta y^\cCart_{\max} \cdot y^\cLSP_1 - \frac{\Delta y^\cCart}{2}
    .
    \label{eq-yproj2}
\end{align}
The last term in \cref{eq-yproj2} is needed to center the panel around $\theta^\cLS$ (\cref{fig-projectiony}, panels A and B).

Finally, for the panel placement, we operate in polar coordinates and offset the azimuthal angle with the correct value $\theta^\cLSP$.
\begin{align}
    \begin{split}
        r_3^\cPol
            &= \sqrt{\left(x^\cLSP_1\right)^2 + \left(y^\cLSP_2\right)^2}
        ,
        \\
        \theta_3^\cPol
            &= \arctan\left(\frac{y^\cLSP_2}{x^\cLSP_1}\right) + \theta^\cLSP
        .
    \end{split}
    \label{eq-projypolar}
\end{align}
Once again, an offset of $\pi$ might be needed depending on the implementation of $\arctan$.
Converting back to Cartesian coordinates gives us the values that can be plotted (\cref{fig-projectiony}, panel C):
\begin{align}
    \begin{split}
        x^\cLSP &= r_3^\cPol \cdot \cos(\theta_3^\cPol), \\
        y^\cLSP &= r_3^\cPol \cdot \sin(\theta_3^\cPol)
        \label{eq-projyresult}
    \end{split}
    .
\end{align}
Alternatively to \cref{eq-projypolar} one could remain in Cartesian space and use a rotation matrix to achieve the same result.

\subsubsection{Adding context from the original data}
\label{sec-addingcontext}
Based on the previous sections all values that are displayed using \gls{lstein} are constrained to the $\ell_\infty$ unit ball.
To add the physical context originally recorded by the inputs ($x^\cCart$, $y^\cCart$, $\theta^\cCart$) back to the visualization, we adjust the axis ticks accordingly.
This means, that we also apply \cref{eq-minmaxscale} to all the axis ticks provided by the user to convert their locations to the $\cLSP$ coordinate system.
Then it is just a matter of under-plotting them as lines and arcs behind the projected data from \cref{sec-projecttheta,sec-projecty}.
This concludes our final transformation: $\cLSP \mapsto \cLS$.
We show the final result and a visualization of the \gls{lstein} grid and axis ticks in \cref{fig-lsteinvars}.

\subsection{Note on projection effects}
\label{sec-projectioneffects}
Comparing \cref{fig-projectiontheta,fig-projectiony,fig-lsteinvars} one can immediately see that the two methods do not yield the same results.
This is intended and has to be taken into account when plotting your data.
The slight distortion in \cref{fig-projectiontheta} (panel A of \cref{fig-lsteinvars}) is not present in \cref{fig-projectiony} (panel B of \cref{fig-lsteinvars}).
Note, however, that the distortion almost guarantees that the data is contained within the panel, while values when using \texttt{y\_projection\_method="y"} might overflow at the borders (i.e., panel B in \cref{fig-lsteinvars}).
We do not recommend drawing conclusions from \gls{lstein} exclusively, but rather combine it with other visualization techniques.

In cases where the linearity of trends in the series are crucial for the analysis we recommend using \texttt{y\_projection\_method=y}.
The same is true if the morphology of each individual series is of importance for the science case, since \texttt{y\_projection\_method=y} reduces the projection effects.
If, on the other hand, only intra-series morphology differences are of interest, \texttt{y\_projection\_method=theta} will be advantageous because all plotted data-points are on the grid (no overflow).
For web-applications, where one wants to avoid data-overflow, we recommend using \texttt{y\_projection\_method=theta}.

\section{Extension to other fields}
\label{sec-extensions}
We designed \gls{lstein} to be general enough for a variety of applications.
In principle, \gls{lstein} can be applied to any 2.5D dataset, no matter the field of research or industry.
This section will provide a non-exhaustive list of contexts where \gls{lstein} might be useful.

\subsection{Spectral evolution over time}
\label{sec-spectraovertime}

Another mode of observation in astronomy is spectroscopy, where one splits the incoming light into its independent wavelength components (essentially an infinite amount of passbands).
Spectroscopy usually has a very high spectral (wavelength) resolution, but is sparse in time.
The evolution of a spectrum over time does, however, encode important information about astrophysical transients such as \acrlongpl{sn} \citep[i.e.,][]{Filippenko1997_SNTypes}.

\begin{figure}[t]
    \centering
    \includegraphics[width=\linewidth]{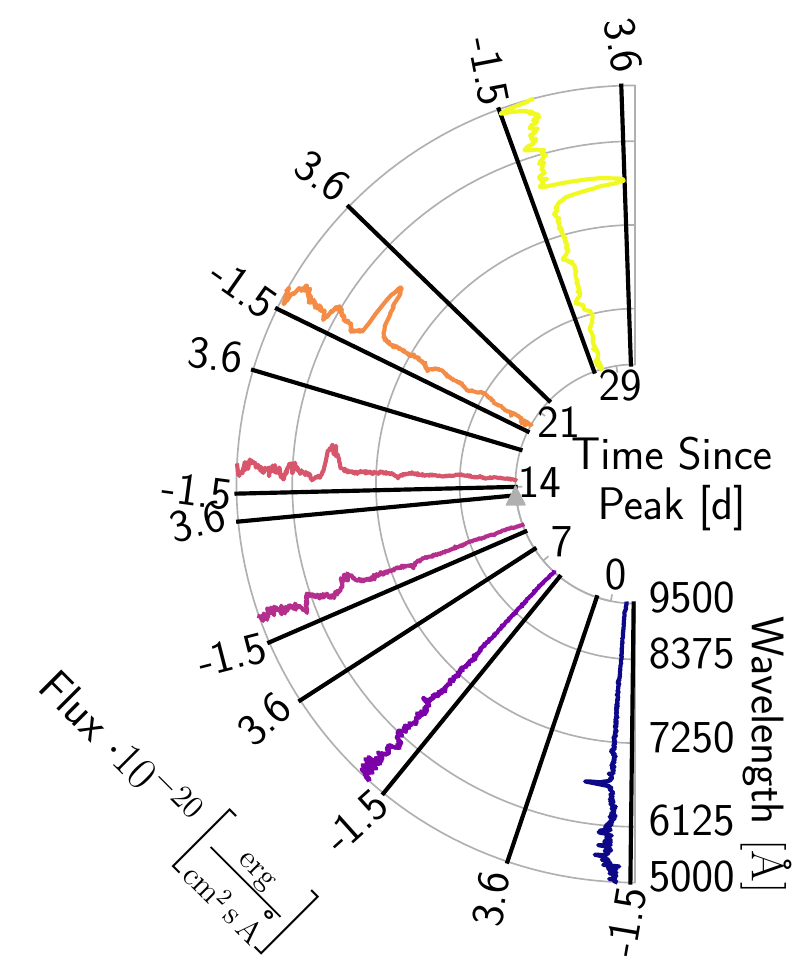}
    \caption{
        Example application showing temporal evolution of spectra from a \acrlong{sn} explosion.
        The displayed object is SN2023ixf.
        Displayed spectra are from the \acrlong{desi} \citep[\acrshort{desi}][]{Levi2019_DESI}.
        Note, that in order to reduce distorting projection effects, we rescaled the wavelength to the range $[0,10]$ (similar range to flux values) reapplied the physical tick-labels afterward.
        The alternative would be to use \texttt{y\_projection\_method="y"}.
    }
    \label{fig-spectralevolution}
\end{figure}

One could use \gls{lstein} to plot the spectral evolution over time by using the following mapping:
\begin{align}
    \begin{split}
        \lambda &\mapsto x^\cLS \\
        f &\mapsto y^\cLS \\
        t &\mapsto \theta^\cLS \\
    \end{split}
    ,
    \label{eq:spectrummap}
\end{align}
where $\lambda$ denotes wavelength, $f$ is the recorded flux, and $t$ is time.
We show an example in \cref{fig-spectralevolution}, where it is visible which measurements have been taken closer together or further apart.

\subsection{Radio Astronomy}
\label{sec-glitchesradioastronomt}
Pulsar timing research, a field in radio-astronomy, is concerned with the dispersion of a pulse over different frequencies.
Traditionally, this analysis is done by plotting a heatmap of intensity $I$ in dependence of pulsar-phase $\phi$ and frequency band $f$.
Heatmaps make it challenging to discern small-scale variations by eye because they are based on color-variation \citep[i.e.,][]{Moreland2009_DivergingColormaps,Moreland2015_Colormaps}.
This makes identification of radio-glitches (inconsistencies in the data that can influence interpretation) difficult, especially since the pulse-intensity dominates the color-scale.
Additionally, radio-astronomy is highly influenced by modern society due to the extensive use of radio-frequencies for communications, which contaminate specific frequency bands to the point where they might become unusable.

With \gls{lstein}, one could map as follows:
\begin{align}
    \begin{split}
        \phi &\mapsto x^\cLS \\
        I &\mapsto y^\cLS \\
        f &\mapsto \theta^\cLS \\
    \end{split}
    .
    \label{eq-radiomapping}
\end{align}
This way, contaminated bands can simply be removed, because the position in frequency space is preserved.
Additionally, showing the actual vertical values simplifies identification of radio-glitches and small-scale features.
To ease the glitch-discovery even further, one can exploit the projection towards the edges of the \texttt{LSteinCanvas} and amplify regions of interest.

\begin{figure}[t]
    \centering
    \includegraphics[width=0.95\linewidth]{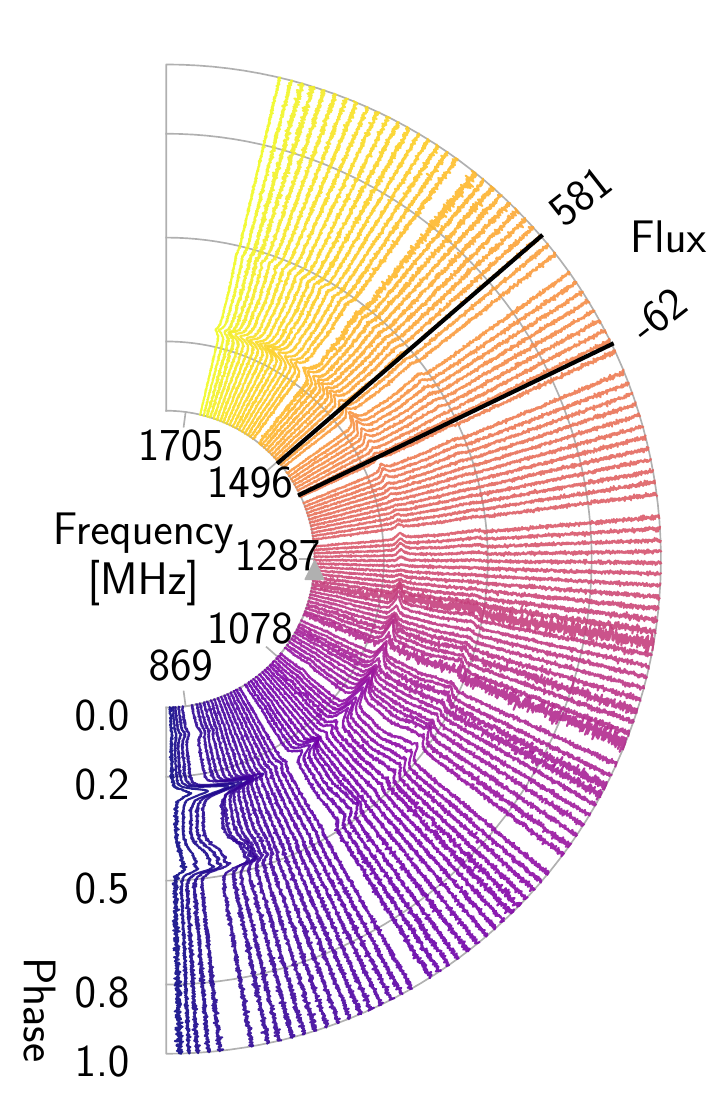}
    \caption{
        Example application to pulsar timing research.
        The displayed pulsar is J2145-0750 with observations from 2023-03-07-09:08:21.
        We simulate randomly missing frequency bands in the data.
        Because the dataset is very dense in $\theta$ the panels do overlap.
        To preserve readability we only indicate a single panel to give the reader an idea of panel sizes.
        Note that we used \texttt{y\_projection\_method="y"} for this example because only one series was plotted per panel and large values lead to strong projection effects when applying \texttt{y\_projection\_method="theta"}.
        Flux is provided in arbitrary units.
    }
    \label{fig-pulsartiming}
\end{figure}

We show an example for an application to a pulsar observed with \gls{meerkat} \citep{Booth2009_MeerKAT,Johnston2020_MeerKAT,Bailes2020_MeerKAT} in \cref{fig-pulsartiming}.
The data was obtained from the Pulsar Portal\footnote{Homepage: \url{https://pulsars.org.au/}} (Bailes et al., in prep).
One can see the pulse profile, and location of missing, as well as highly contaminated radio frequency bands.
This particular example did not show features that migrate across the phase-dimension but in other examples we explored also these were identifiable.

\subsection{Learning curves}
\label{sec-learngcurves}

\begin{figure}[t]
    \centering
    \includegraphics[width=\linewidth]{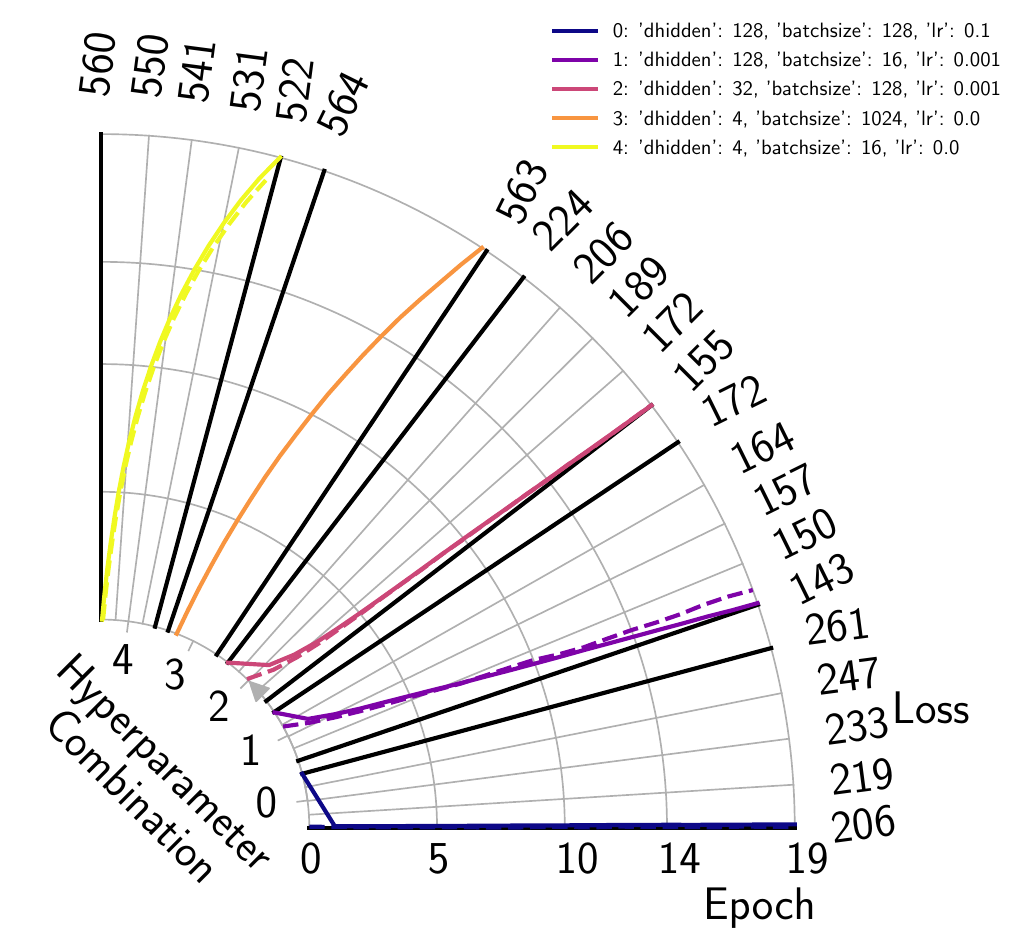}
    \caption{
        Example applying \gls{lstein} to visualize a hyperparameter search.
        Solid lines denote training loss, dashed lines validation loss.
    }
    \label{fig-loss}
\end{figure}

In \gls{ml}, learning-curves are commonly used to quantify a model's performance over time.
A straightforward way of using \gls{lstein} in this context would be to map as follows:
\begin{align}
    \begin{split}
        \text{epoch} &\mapsto x^\cLS \\
        \mathcal{L} &\mapsto y^\cLS \\
        \mathcal{H} &\mapsto \theta^\cLS \\
    \end{split}
    ,
    \label{eq:mlmap}
\end{align}
where $\mathcal{H}$ is representative for any hyperparameter (i.e., number of layers in a neural network, regularization parameter, number of neurons).
The model's performance (loss) is denoted as $\mathcal{L}$.
An example of \gls{ml} learning curves can be found in \cref{fig-loss}.
The curves in \cref{fig-loss} are from training an \acrlong{autoencoder} \citep{Kramer1991_Autoencoder} to encode \gls{mnist} handwritten digits \citep{Deng2012_Mnist}\footnote{See \url{https://github.com/TheRedElement/LStein/blob/paper/data/auto_encoder_hypsearch.ipynb} for the training script.}.

\subsection{Spiking neurons}
\label{sec-spikingneurons}
\begin{figure}[t]
    \centering
    \includegraphics[width=\linewidth]{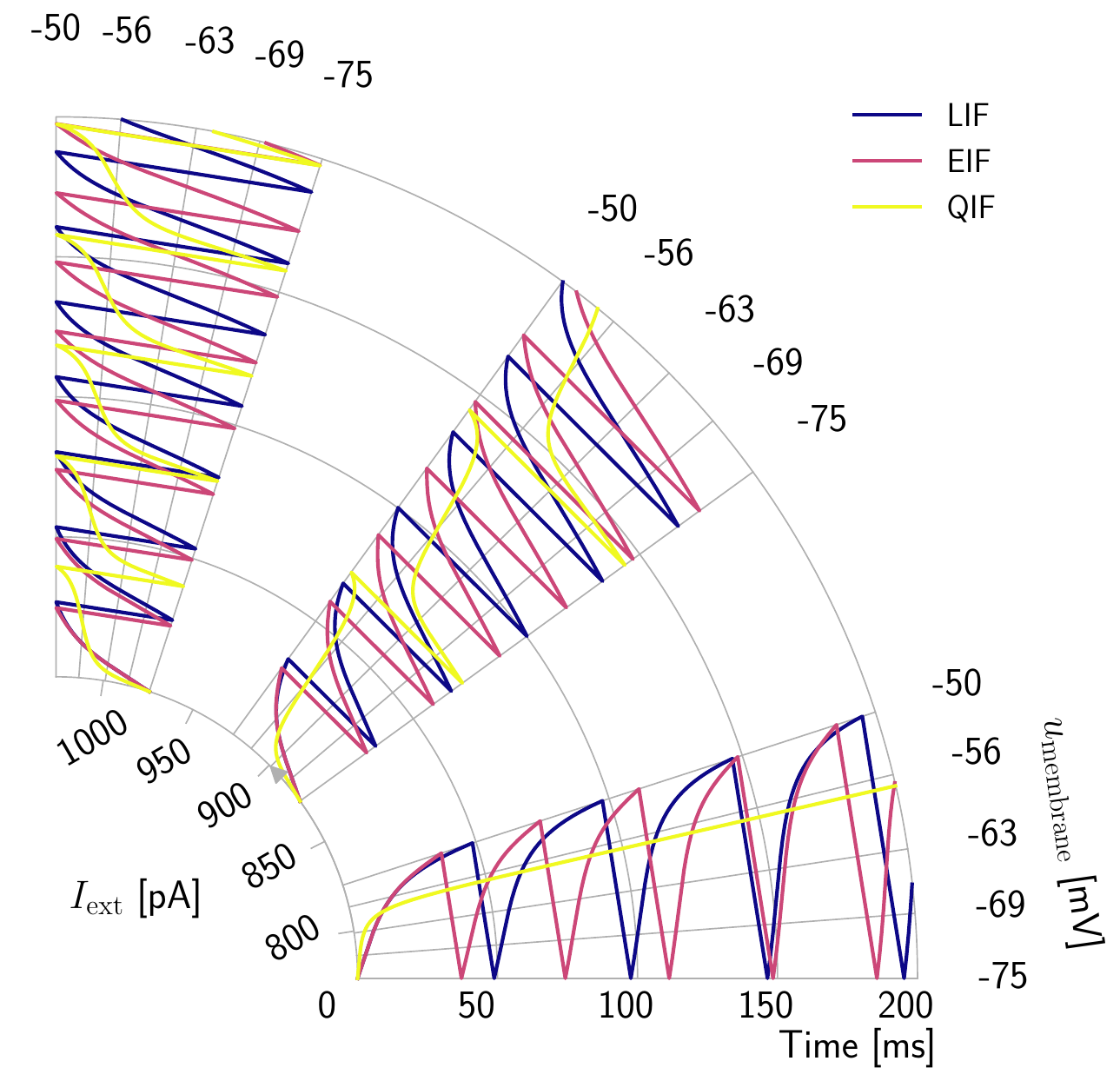}
    \caption{
        Example applying \gls{lstein} to \glspl{snn}.
        Different colors denote different neuron models (Leaky Integrate and Fire -- LIF; Exponential Integrate and Fire -- EIF; Quadratic Integrate and Fire -- QIF).
        Simulations have been done with \gls{brianii} \citep{Stimberg2019_BRIAN2}.
    }
    \label{fig-snn}
\end{figure}

\Acrlongpl{snn} \citep[\acrshortpl{snn},][]{Maass1997_SNN} are biologically inspired neural networks that use discrete spikes to propagate information.
These networks are extensively studied in computational neuroscience and are especially interesting because they resemble the biological inner workings of a brain much closer than the \glspl{ann} used in \gls{dl}.
\Gls{lstein} can facilitate an easy comparison of several neuron-types used in a \gls{snn} with different (constant) input current $I_\mathrm{ext}$.
To do so one would map like so:
\begin{align}
    \begin{split}
        t &\mapsto x^\cLS \\
        u_\mathrm{membrane} &\mapsto y^\cLS \\
        I_\mathrm{ext} &\mapsto \theta^\cLS \\
    \end{split}
    ,
    \label{eq:snnmap}
\end{align}
with $u_\mathrm{membrane}$ being the neurons membrane potential.
In this setting it would be easy to plot several different neuron-types in each panel (see \cref{fig-snn}).
One can take advantage of the fact that color is not used in the encoding process to plot lines of the same neuron type in the same color.
This way, it is easy to spot the same neuron type and compare across panels.
Other quantities that might be interesting when displayed as $\theta$ are: the number of pre-synaptic neurons, or concentrations of chemical tracers.

\section{Known issues and workarounds}
\label{sec-knownissuesandworkarounds}

\subsection{Error bars}
\label{sec-errorbars}
\begin{figure}[t]
    \centering
    \includegraphics[width=\linewidth]{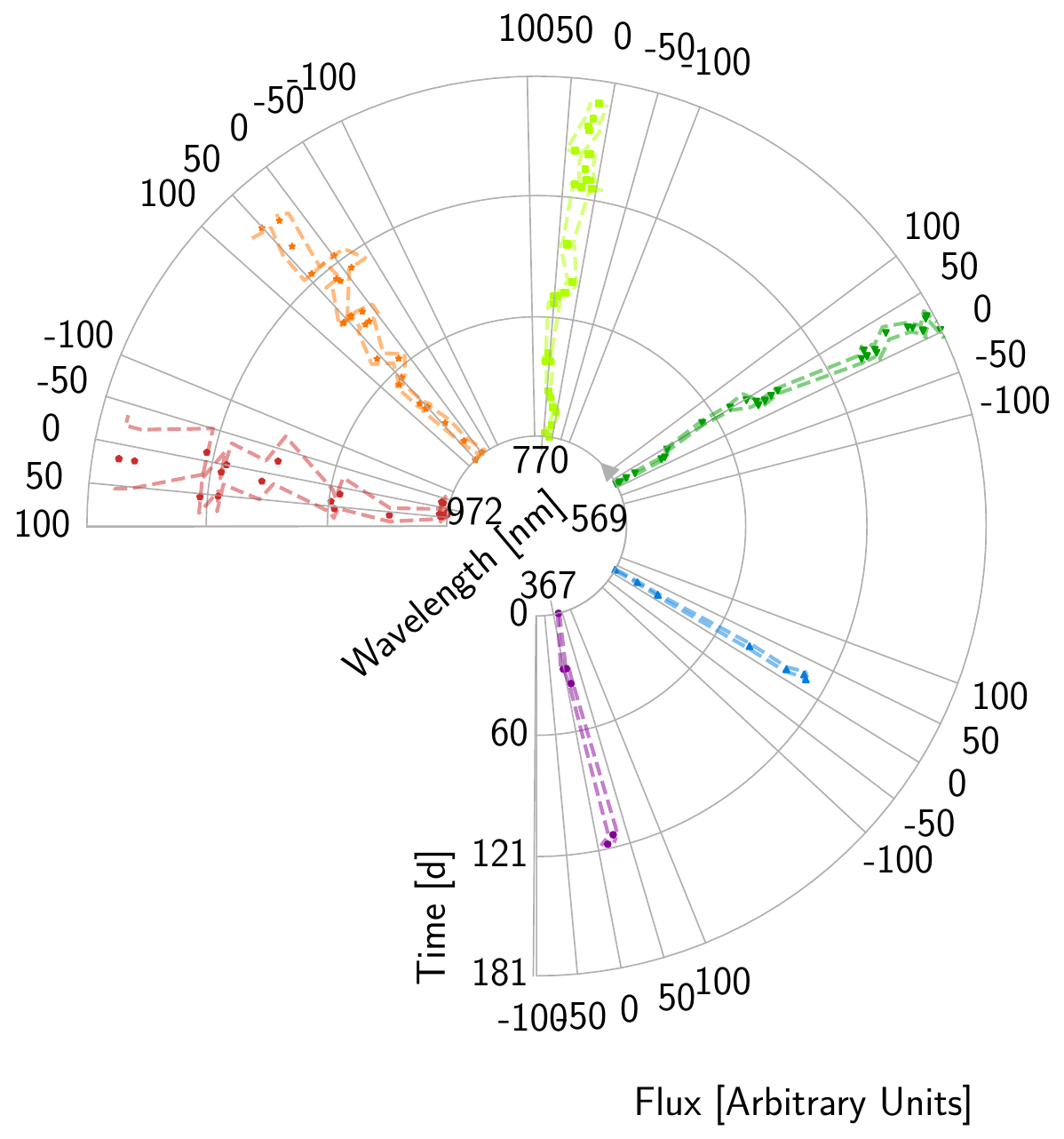}
    \caption{
        Workaround for showing error bars.
        The scatter are the actual measurements, dashed lines denote error regions.
    }
    \label{fig-errorregions}
\end{figure}

In the current implementation of \gls{lstein}, we do not support the plotting of error bars.
The reason is, that these are affected by the projection, which makes it nontrivial to plot them in a consistent manner.
An easy workaround is to plot error regions in the form of two additional lines $y\,\pm\,\vert\Delta y\vert$ (see \cref{fig-errorregions}).
The uncertainty in $y$ is hereby denoted as $\Delta y$.
Note, that the projection effects have to be taken into account here as well, which means that one has to allocate additional buffer in the y-limits to make sure error regions are not cropped at the edge of the \texttt{LSteinPanel}.

Additionally, error regions are already commonly used in scientific plotting.
They are easy to customize and encode the same information as error bars.
As an alternative to error regions one could also filter \glspl{lc} based on the signal to noise ratio before plotting, thus, implicitly encoding the error bar information.
In that context it is also important to keep in mind that \gls{lstein} is a visualization technique complementing existing methods that support error bar visualization.

\subsection{Series types}
\label{sec-seriestypes}

For similar reasons to \cref{sec-errorbars} we only support scatter- (\texttt{seriestype="scatter"}) and line-plots (\texttt{seriestype="line"}) at the moment.
There is no clean workaround as of now, but we encourage users to open a pull-request on \gls{github}\footnote{\url{https://github.com/TheRedElement/LStein/pulls}}, in case they implement other series types.
It has yet to be explored, if there are other series types besides scatter- and line-plots where \gls{lstein} could be a useful alternative for visualization.

\subsection{Inverting \texttt{thetaticks}}
\label{sec-invertingthetaticks}
When passing a reverse sorted (descending order) list as \texttt{thetaticks} to \texttt{LSteinCanvas}, \gls{lstein} will not adjust the displayed tick labels correctly.
Instead, it will pair the passed argument (sorted in descending order as tick labels) with a rescaled version (sorted in ascending order) used for the tick-positions.
This can be fixed by passing explicit tick labels to \texttt{LSteinCanvas}: \texttt{(thetaticks\_descending,thetaticks\_ascending)}.
The reverse sorted list is \texttt{thetaticks\_descending}, the labels to be used are \texttt{thetaticks\_ascending}.

\subsection{Log scale}
\label{sec-logscale}
There is no dedicated implementation of a $\log$-scale because this can internally lead to unpredictable projection effects.
However, one can easily achieve the same effect by manually scaling the data and exploiting the customization options.
Specifically, one can set custom ticks by means of the \texttt{xticks} keyword argument (equivalently for $\theta^\cLS$ and $y^\cLS$).
We provide an example in \cref{lst-logscale}.

\begin{lstlisting}[
    float,    
    caption={
        Example skeleton for applying a $\log$-scale.
    },
    label={lst-logscale},
    style=vscodedarkstyle,
]
x = ...
y = ...
theta = ...

x = [np.log10(xi) for xi in x]
y = [np.log10(yi) for yi in y]
theta = [np.log10(thi) for thi in theta]
xticks = np.logspace(0,5,6)
yticks = np.logspace(0,5,6)
thetaticks = np.logspace(0,5,6)
LSC = LSteinCanvas(
    thetaticks, xticks=xticks,
    yticks=yticks, ...
)
\end{lstlisting}

\subsection{Backend subtleties}
\label{sec-backendsubtleties}
\Gls{lstein} was mainly implemented for \gls{pymatplotlib}.
Therefore, all the internal customization arguments follow the \gls{pymatplotlib} names.
For the \gls{plotly} backend, we wrote a translator that translates some of the arguments to the native names.
However, as translation is not trivial, we do not guarantee that all arguments have a valid counterpart in \gls{plotly}.
Especially for annotations (axis-labels, tick-labels, etc.), this can lead to issues.
The two workarounds are as follows:
\begin{enumerateinline}
    \item Remove the parameters and add custom annotations instead;
    \item Use the default parameters for these elements.
\end{enumerateinline}

\section{Further development}
\label{sec-furtherdevelopment}

The modular nature of \gls{lstein} enables building extensions to the existing package with relative ease.
Future work will aim to include errorbars and a panel-span encoding into \gls{lstein}.
Applications where encoding information into the panel-span could be useful, include the following:
\begin{itemize}
    \item Requirement of the knowledge of the passband width in addition to its central wavelength.
    \item Encoding of integration time when taking a specific spectrum (\cref{sec-spectraovertime}).
    \item Semi-random input current with some characteristic width around an average value (\cref{sec-spikingneurons}).
\end{itemize}
We decided against the implementation of such a feature in our first version of \gls{lstein}, because it would interfere with the comparability of the $y$-ranges across \texttt{LSteinPanel}s.
Furthermore, panels are at higher risk to overlap and the plot could contain too much information for the viewer to process simultaneously.

\section{Summary}
\label{sec-summary}

We presented \gls{lstein} a new plotting framework to visualize 2.5-dimensional (3D where one dimension is sparse) data in 2 dimensions.
The projection is achieved by executing scaling operations in Cartesian and polar coordinates in different steps to ensure preservation of qualitative data-series attributes (\cref{sec-dataseriesprojection}).
Our approach is a first instance to tackle the reduction of wavelength information when plotting \acrlongpl{lc} in multiple passbands.

Comparing our method to traditional approaches shows advantages of using this alternative way of inspecting the data.
Most information can, of course, be extracted from data by looking at it from different angles i.e., combining traditional methods with \gls{lstein}.

We outlined our design principles which include ease of use and customizability.
We follow the application flows of widely used visualization libraries to ensure a shallow learning curve and provide implementations for two popular plotting libraries (\gls{pymatplotlib} and \gls{plotly}).
To make \gls{lstein} even more accessible, we plan to include it on the \gls{fink} broker portal.

To demonstrate that our approach is not limited to astronomy, we provided a non-exhaustive list of examples in which it could also be useful.
Users are encouraged to come up with more creative ways of using \gls{lstein}.

The code is open source and publicly available on \gls{github}\footnote{\gls{github}: \url{https://github.com/TheRedElement/LStein}}.
The \href{https://lstein.readthedocs.io/en/latest/}{documentation} can be accessed through \gls{readthedocs}\footnote{Documentation: \url{https://lstein.readthedocs.io/en/latest/}}.
Finally, we provide a set of known issues and current workarounds.
We plan to mitigate some of those issues in the near future.

We expect that \gls{lstein} will be a useful tool to visualize data from modern astrophysical surveys and hope that also other fields will find value in this approach.

\section*{Acknowledgments}

This work was developed within the \gls{fink} community and made use of the \gls{fink} community broker resources.
\gls{fink} is supported by LSST-France and CNRS/IN2P3.
This research was supported by grants from the Australian Research Council.
AM is supported by the ARC Discovery Early Career Research Award (DE230100055).
Parts of this research were conducted by the Australian Research Council Centre of Excellence for Gravitational Wave Discovery (OzGrav), through project CE230100016.

\section*{Data availability statement}
All data used in this research is publicly available and referenced accordingly in the manuscript.

\section*{Author contribution}
LS:
    Conceptualization,
    Formal analysis,
    Methodology,
    Software,
    Visualization,
    Writing -- review and editing,
    Writing -- original draft.
AM:
    Methodology,
    Supervision,
    Writing -- review and editing.
CJF:
    Methodology,
    Supervision,
    Writing -- review and editing.

\printglossaries

\bibliographystyle{elsarticle-harv}
\bibliography{bib-refs}

\appendix
\section{Arbitrary backends}
\label{sec-additionalcodeexamples}

We show the most basic recipe to use \gls{lstein} for production of graphical output in your favorite backend in \cref{lst-lsteinrecipe}.
Note that you might have to make minor modifications for some plotting frameworks.
\begin{figure*}
    \begin{lstlisting}[
        caption={Recipe to show \gls{lstein} as a figure in an arbitrary backend.},
        label={lst-lsteinrecipe},
        style=vscodedarkstyle,
        ]
        #import package
        from lstein import lstein
        from lstein.utils import polar2carth

        #setup a new canvas
        LSC = lstein.LSteinCanvas(...)
        #add artists (will also generate panels)
        LSC.plot(thetavals, xvals, yvals, seriestype="scatter", ...)
        LSC.plot(thetavals, xvals, yvals, seriestype="plot", ...)

        ls_xaxis = LSC.compute_xaxis()          #get quantities defining the x-axis
        ls_thetaaxis = LSC.compute_thetaaxis()  #get quantities defining the theta-axis
        ls_ylabel = LSC.compute_ylabel()        #get position of y-label

        ... #plot elements in your favourite backend

        #draw panels
        for LSP in LSC.Panels:
            #get panel boundaries
            theta_offset, theta_lb, theta_ub = LSP.get_thetabounds()
            r_lb, r_ub = LSP.get_rbounds()
            r_bounds = np.array([r_lb, r_ub])

            #get yticks
            ytickpos_th, yticklabs = LSP.get_yticks(theta_lb, theta_ub)

            #convert to cartesian for plotting
            x_lb, y_lb  = polar2carth(r_bounds, theta_lb)
            x_ub, y_ub  = polar2carth(r_bounds, theta_ub)
            x_bounds = np.array([x_lb,x_ub])
            y_bounds = np.array([y_lb,y_ub])

            pad = LSP.yticklabelkwargs["pad"]   #padding for yticklabels
            r_, th_ = np.meshgrid(r_bounds, ytickpos_th)
            ytickpos_x, ytickpos_y           = polar2carth(r_, th_)
            yticklabelpos_x, yticklabelpos_y = polar2carth((1+pad)*r_ub, ytickpos_th)

            ... #plot elements in your favourite backend

            for ds in LSP.dateseries:
                #plot series in your favourite backend
                if ds["seriestype"] == "plot":
                    plot(ds["x"], ds["y"])
                if ds["seriestype"] == "scatter":
                    scatter(ds["x"], ds["y"])
    \end{lstlisting}
\end{figure*}

\section{Loading a \gls{json} file in \gls{javascript}}
A \gls{javascript} function to an \gls{lstein} plot saved as \gls{json} file within a website is provided in \cref{lst-plotfromjson}.
To include a plot on a website using the function in \cref{lst-plotfromjson}, one simply adds an element \texttt{<div class="plotly-jsonplot" data-path="path/to/plot.json"></div>} to the website.
\begin{figure*}
    \begin{lstlisting}[
        caption={Recipe to show \gls{lstein} as a figure in an arbitrary backend.},
        label={lst-plotfromjson},
        language={JavaScript},
        ]
        async function plotFromJson() {
            try {
                //get all relevant divs (divs to plot into)
                const plyDivs = document.getElementsByClassName("plotly-jsonplot");

                //deal with all relevant divs
                for (const element of plyDivs) {

                    //get path to plotting data
                    const dataPath = element.dataset.path;

                    //fetch json file
                    const response = await fetch(dataPath);

                    //init figure
                    const fig = await response.json();

                    //adjust layout (no padding)
                    fig["layout"]["margin"] = {};
                    fig["layout"]["margin"]["t"] = 0;
                    fig["layout"]["margin"]["l"] = 0;
                    fig["layout"]["margin"]["r"] = 0;
                    fig["layout"]["margin"]["b"] = 0;

                    //show the figure
                    Plotly.newPlot(element, fig.data, fig.layout, fig.config).then(
                        () => {
                            Plotly.addFrames(element, fig.frames);
                        }
                    );
                }

            } catch (error) {
                console.error("Error fetching or plotting data:", error);
            }
        }
        plotFromJson();
        window.addEventListener("resize", function(event) {
            plotFromJson();
        });
    \end{lstlisting}
\end{figure*}

\end{document}